\documentclass{article}

\PassOptionsToPackage{numbers, compress}{natbib}

\usepackage[preprint]{neurips_2024}

\usepackage{packages}

\newcommand{\norm}[1]{\lVert #1 \rVert}

\newcommand{\doubleapex}[1]{``#1''}
\DeclareMathOperator*{\argmax}{argmax}
\DeclareMathOperator*{\argmin}{argmin}
\DeclareMathOperator*{\ewmax}{ewmax}
\DeclareMathOperator*{\ewmin}{ewmin}
\newcommand\ddfrac[2]{\frac{\displaystyle #1}{\displaystyle #2}}
\newcommand\nan{\text{NaN}}
\newcommand\convexhull{\text{CH}}

\theoremstyle{definition} 

\newtheorem{assumption}{Assumption}
\theoremstyle{remark}
\newtheorem{remark}{Remark}

\makeatletter
\newcommand{\RemoveAlgoNumber}{\renewcommand{\fnum@algocf}{\AlCapSty{\AlCapFnt\algorithmcfname}}}
\newcommand{\RevertAlgoNumber}{\algocf@resetfnum}
\makeatother

\title{Distributed clustering in partially overlapping feature spaces}

\author{%
  Alessio Maritan\\
  University of Padova\\
  \And
  Luca Schenato\\
  University of Padova\\
}

\begin{document}

\maketitle

\begin{abstract}
We introduce and address a novel distributed clustering problem where each participant has a private dataset containing only a subset of all available features, and some features are included in multiple datasets. 
This scenario occurs in many real-world applications, such as in healthcare, where different institutions have complementary data on similar patients. We propose two different algorithms suitable for solving distributed clustering problems that exhibit this type of feature space heterogeneity. The first is a federated algorithm in which participants collaboratively update a set of global centroids. The second is a one-shot algorithm in which participants share a statistical parametrization of their local clusters with the central server, who generates and merges synthetic proxy datasets. In both cases, participants perform local clustering using algorithms of their choice, which provides flexibility and personalized computational costs.
Pretending that local datasets result from splitting and masking an initial centralized dataset, we identify some conditions under which the proposed algorithms are expected to converge to the optimal centralized solution. Finally, we test the practical performance of the algorithms on three public datasets.
\end{abstract}

\section{Introduction}

Clustering, the task of grouping unlabeled data points based on their similarity, is a fundamental pillar of data mining and machine learning, widely applied across diverse fields such as pattern recognition, image processing, bioinformatics, and customer segmentation. Traditional centralized clustering algorithms, which assume that all data are stored in a single location, face several limitations in contemporary data landscapes, characterized by massive data volumes and naturally distributed data islands.
In such scenarios, gathering and processing all data in single machine becomes impractical or even impossible due to prohibitive communication costs, computational bottlenecks and especially privacy concerns.
In fact, local datasets typically contain sensitive information that cannot be shared directly with other agents or a central server due to stringent privacy regulations, competitive interests, or individual anonymity preferences.
This has led to a shift towards distributed clustering methodologies, which allow participants to contribute to a global clustering solution without raw data sharing, ensuring that their private information remains confidential.
This collaborative but privacy-preserving approach finds several applications, such as healthcare data analysis, where patient records are distributed across hospitals, and user profiling, where different institutions hold complementary datasets.

In distributed algorithms, iterative communication between participants can quickly become a bottleneck, especially in large-scale systems or environments with limited bandwidth. Therefore, designing clustering algorithms that minimize communication rounds and data exchange is critical for scalability, network performance, and in cases of intermittent connectivity.

\begin{figure}[htb]
    \centering
    \includegraphics[width=0.5\linewidth]{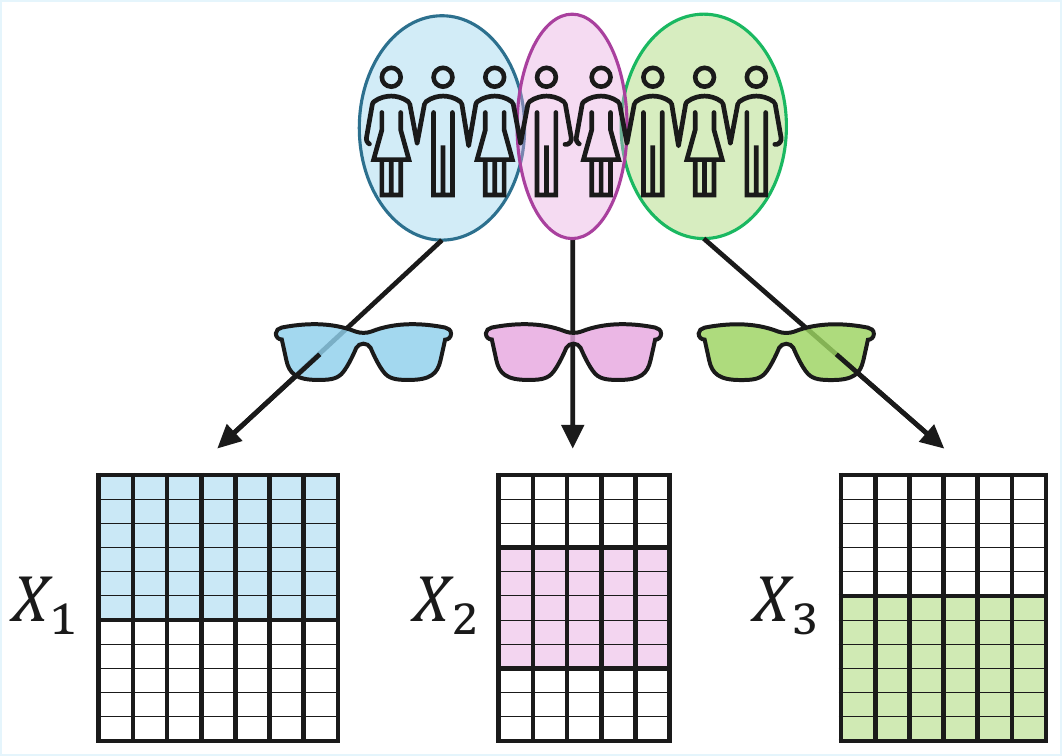}
    \caption{We consider the scenario in which participants collect similar data, but each participant observes only a fixed subset of features and the sets of features observed by different participants partially overlap. In the figure, $X_1, X_2, X_3$ are the local datasets of three participants, where each column is a data point and the features observed by the participants are indicated using colors.}
    \label{fig:intro}
\end{figure}

Differently from most existing work, this paper addresses the distributed clustering scenario in which the feature spaces of the participants only partially overlap.
Equivalently, in the larger feature space given by the union of the feature spaces of the participants, each participant is associated with a mask that allows him to observe only a fixed subset of features.
This scenario is motivated by the fact that, in many practical contexts, individual participants observe partial realizations of some common underlying random variables or phenomena, as shown in Figure \ref{fig:intro}. For example, two clinics may collect different sets of blood markers from patients with similar demographics or medical conditions. Similarly, in sensor networks, individual sensors may record different subsets of environmental parameters due to hardware limitations or specific monitoring objectives.
When performing learning or data mining tasks in these scenarios, the local view of each participant is incomplete, but collectively participants hold a more comprehensive picture. Distributed clustering is a promising tool to combine partial observations and allow participants to converge to better global solutions.
However, there is a gap in the literature about distributed clustering algorithms, since almost no previous work explicitly considers the case where local datasets contain different features and some features are included in multiple datasets. 
In this paper, we formalize the above problem and address it by designing and analyzing two novel algorithms that offer several desirable properties.

Our contributions can be summarized as follows.
\begin{itemize}
    \item We formalize a novel distributed clustering problem in which the local data of different participants belong to different but partially overlapping feature spaces. Tho goal is to recover the optimal centralized solution without raw data sharing.
    \item Assuming a star network topology, where all participants are connected to a central server, we propose two algorithms to solve the above problem. The first is a federated clustering algorithm that alternates between local clustering and global aggregation of local centroids. The second is a one-shot distributed algorithm consisting of two phases. In the first phase, each participant runs an arbitrary clustering algorithm on his private dataset and fits a probability distribution on each cluster. In the second phase, the central server generates a synthetic dataset from these probability distributions and aggregates the local clustering solutions. In both cases, participants can perform local clustering using different algorithms, so as to adapt the computational cost to their computing resources.
    \item By pretending that the private datasets of participants result from splitting and masking a common central dataset, we provide theoretical arguments in support of the proposed algorithms. In particular, we identify some sets of feasibility conditions under which our algorithms are expected to converge to the optimal centralized solution.
    \item We provide numerical experiments showing the effectiveness of the proposed algorithms, considering different public datasets, data distributions and overlaps between feature spaces.
\end{itemize} 

The remainder of this paper is organized as follows. Section \ref{sec:related_work} reviews related work, focusing on clustering with missing values and distributed clustering. Section \ref{sec:problem_formulation} formally defines the problem of distributed clustering of private data in partially overlapping feature spaces. Sections \ref{sec:federated_algorithm} introduces the federated algorithm, and Section \ref{sec:one_shot_algorithm} presents the one-shot algorithm. Section  \ref{sec:experiments} shows experimental results demonstrating the effectiveness of the proposed algorithms. Finally, Section \ref{sec:conclusions} concludes the paper and discusses future research directions.
\section{Related work} \label{sec:related_work}

The first papers on clustering date back almost 70 years, and over these decades a myriad of centralized, parallel, and distributed algorithms have been proposed. However, there is no distributed or parallel clustering algorithm addressing the case where the feature spaces of participants are different and only partially overlap. In this section we focus on topics related to our specific setting, such as clustering in presence of missing values and vertical federated clustering.

Real world datasets are often affected by missing values, where the underlying missingness mechanism can be completely random, random or not random \cite{rubin1976inference}. In our case, the fact that participants observe only a fixed subset of features can be seen as a special type of not random missingness. Fragmented data can be handled using marginalization, which consists in removing incomplete data points, or data imputation, which refers to estimating missing values from the rest of the dataset. For example, statistical imputation methods typically fill in missing values with their empirical mean, or with the output of a regression model, or using expectation-maximization. Clustering-based imputation methods involve clustering intact data points and using centroids to fill in missing values. Other options include model-based imputation, matrix completion and generative imputation using deep learning \cite{emmanuel2021survey, fan2020polynomial, sun2023deep}.
A few algorithms, e.g. \cite{wang2019k, zhang2021gaussian}, combine data imputation with clustering and can be directly applied to data with missing values. Other works propose alternative approaches: \cite{chi2016k} modifies $K$-means by minimizing the objective function over the observed entries only, and \cite{wagstaff2004clustering} addresses missing values using soft constraints.

Moving to distributed clustering, \cite{singh2013comparative} reviews distributed versions of classical clustering algorithms. More recent works address federated learning only, in which participants are connected to a central server responsible for orchestration and local information aggregation. The basic working principle of federated learning is that participants receive a model from the central server, train it on their private local data, and send a model update back to the server, such as the gradient with respect to the optimization variables \cite{mcmahan2017communication}. In federated clustering, not to be confused with clustered federated learning, the model to be learned is the clustering solution. For example, different versions of federated $K$-means are proposed in \cite{garst2024federated, zhu2023f3km, zhou2022memory}.
Federated learning has two main variants: horizontal and vertical. Horizontal federated learning is used when participants share the same feature space but have data with different sample identifiers (IDs), while vertical federated learning is used when participants share the same sample IDs but have different feature spaces.
In this paper we consider a third case, where participants have both different sample IDs and feature spaces, and the feature spaces of some participants partially overlap. Since this can be seen as a special type of distributed clustering with missing values, we focus on federated clustering algorithms for incomplete data. Reference \cite{ngo2023federated} addresses federated clustering of longitudinal behavioral trial data, and proposes a multiple imputation fuzzy clustering algorithm. Most other works about federated clustering with missing values \cite{li2025vertical, yan2024federated, chen2023federated} consider multi-view clustering, where multiple feature representations, called views, are available for each data point. In particular, they address multi-view clustering in the vertical federated learning setting, where participants have the same sample IDs and different views. Only a few works consider hybrid federated multi-view settings where participants hold data varying in terms of both views and sample IDs, e.g. \cite{chen2024bridging, ren2024novel}. In these works, it is unclear whether the feature spaces of the different participants overlap, and this potential overlap is not explicitly exploited. Furthermore, these works follow completely different approaches from the one proposed in this paper.
\section{Problem Formulation} \label{sec:problem_formulation}

We address a distributed clustering problem where each of $n$ participants has a local private dataset containing data points in a different feature space. Local datasets are correlated with each other, in the sense that participants observe different realizations of a common data-generating process from different points of view. The feature spaces of the participants partially overlap, allowing meaningful comparisons under suitable assumptions. Considering the feature space resulting from the union of the local feature spaces, we can rewrite the problem by associating each participant with a binary mask that allows him to observe only a fixed subset of features. We consider a star network topology where all participants are connected to a central server that is responsible for aggregating local information. The server is honest-but-curious, and raw data sharing is prohibited.

We focus on hard clustering, the task of partitioning unlabeled data points based on their features, resulting in a set of disjoint groups (called clusters) that exhibit intra-cluster similarity and inter-cluster dissimilarity. The central point that represents a cluster, calculated as the arithmetic mean of all data points within that cluster, is called the centroid. Instead, the actual data point within a cluster that is most centrally positioned, i.e. that minimizes the total dissimilarity to every other point in the cluster, is called a medoid.

\textbf{Notation.} Given a positive integer $n$, we define $[n] = \{1, \dots, n \}$. The vector $e_i$ is the $i$-th vector of the canonical basis in $\mathbb{R}^d$. The binary logic operators AND and OR are $\wedge$ and $\vee$, respectively. For a set of points $S$, the cardinality is $|S|$ and the convex hull is $\convexhull(S)$. The operators $\ewmax(\cdot)$ and $\ewmin(\cdot)$ return the entry-wise maximum and minimum computed over the elements of a set, respectively.
We use the letters $(C,c)$ to denote the clusters and centroids identified by participants, and $(O,o)$ to denote the optimal (or true) clusters and centroids. In general,  subscripts indicate participants, while superscripts indicate clusters. For example, $\mathscr{c}_i = [ c_i^1, c_i^2, \dots]$ is the list of centroids of participant $i$, and $\mathscr{c} = \mathscr{c}_1 \cup \dots \cup \mathscr{c}_n $ contains the centroids from all participants.

\subsection{Data partitioning and masking}

\begin{figure}[htb]
    \centering
    \includegraphics[width=0.5\linewidth]{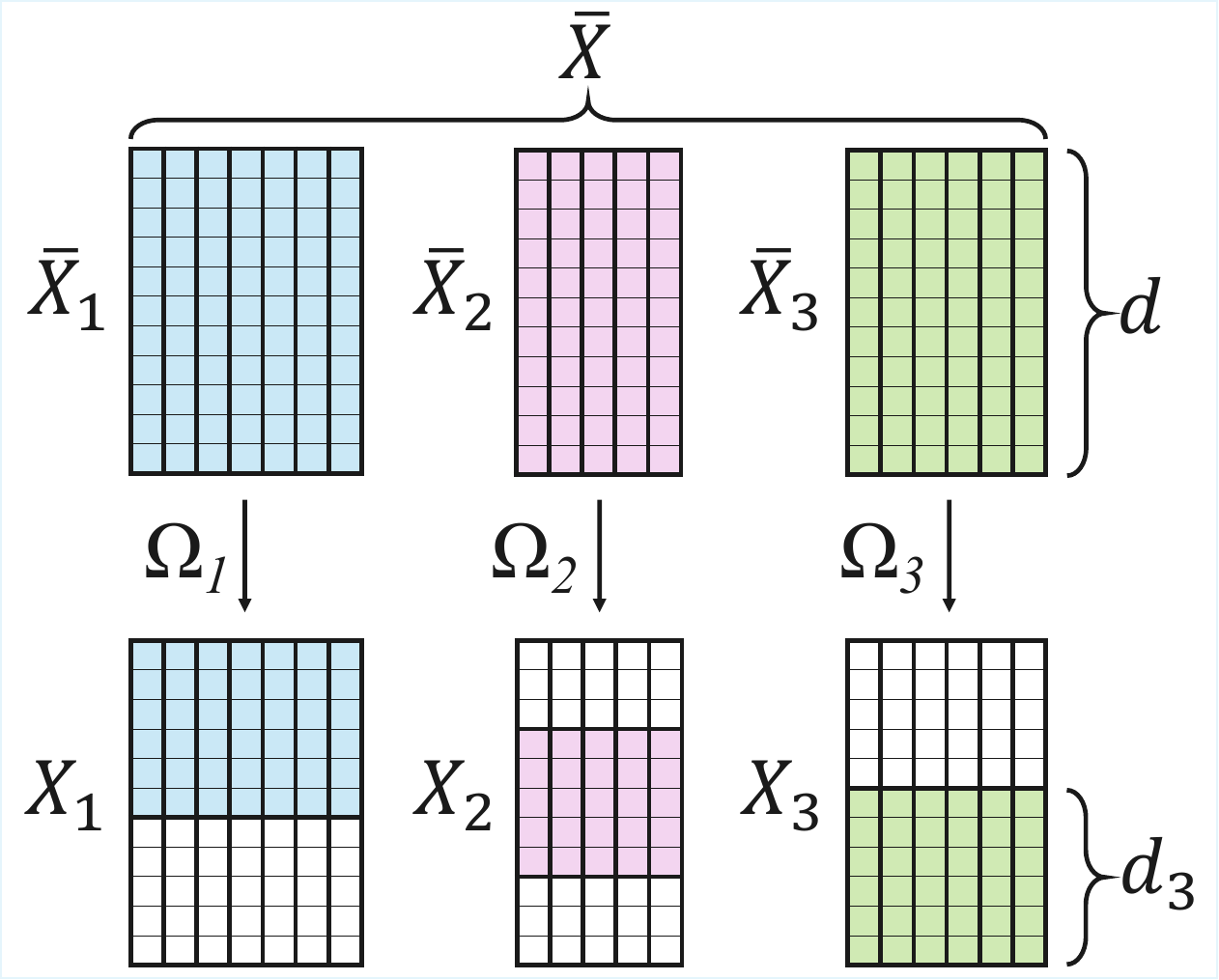}
    \caption{In order to analyze the proposed algorithms, we pretend that the local datasets were generated by splitting a fictitious central dataset $\bar{X}$ and masking each dataset $\bar{X}_i$ with a diagonal, binary, projection matrix $\Omega_i$.}
    \label{fig:partition_and_masking}
\end{figure}

For modeling and analysis purposes, we pretend that local datasets result from the following procedure. Consider a fictitious central dataset $\bar{X}$ containing complete data points in $\mathcal{X}^d$, the $d$-dimensional feature space resulting from the union of the local feature spaces. Note that $\bar{X}$ is introduced only to analyze the proposed algorithms and does not exist in practice. Each participant $i \in [n]$ is associated with a binary, diagonal, projection matrix $\Omega_i \in \{0, 1\}^{d \times d}$ whose diagonal contains $d_i \leq d$ non-zero entries. First, the dataset $\bar{X}$ is partitioned into $n$ disjoint datasets $\{ \bar{X}_1, ..., \bar{X}_n \}$. Partitioning can be heterogeneous, i.e. we also consider the case where local datasets are not i.i.d. and entirely lack observations associated with some ground truth clusters. Then, datasets are masked according to the projection matrices associated with participants, resulting in $\{ X_1, ..., X_n\}$, where each $X_i = \{ \Omega_i x \ | \ x \in \bar{X}_i \}$. This procedure is shown in Figure \ref{fig:partition_and_masking}. The masking matrix $\Omega_i$ sets to zero $d-d_i$ entries of the vector it multiplies. This can be seen as an axis-aligned projection, where the original vector is projected on a lower-dimensional subspace spanned by a subset of the canonical basis. The subspace containing the non-masked coordinates is the actual feature space of participant $i$ and is denoted by $\Omega_i\mathcal{X}^d$.

\begin{remark}
In practice, participants observe only the non-masked entries of their data points, while the masked ones are set to \nan, standing for \doubleapex{not a number}, which indicates an unknown scalar value. When performing local computations, participants only consider the non-masked entries of their local data points, so that the cost of processing a point $x \in X_i$ is proportional to $d_i$. For the ease of exposition, we always consider full $d$-dimensional vectors and use projection matrices that set the masked entries to zero, addressing masking separately.
\end{remark}

\begin{figure}[htb]
    \centering
    \includegraphics[width=0.5\linewidth]{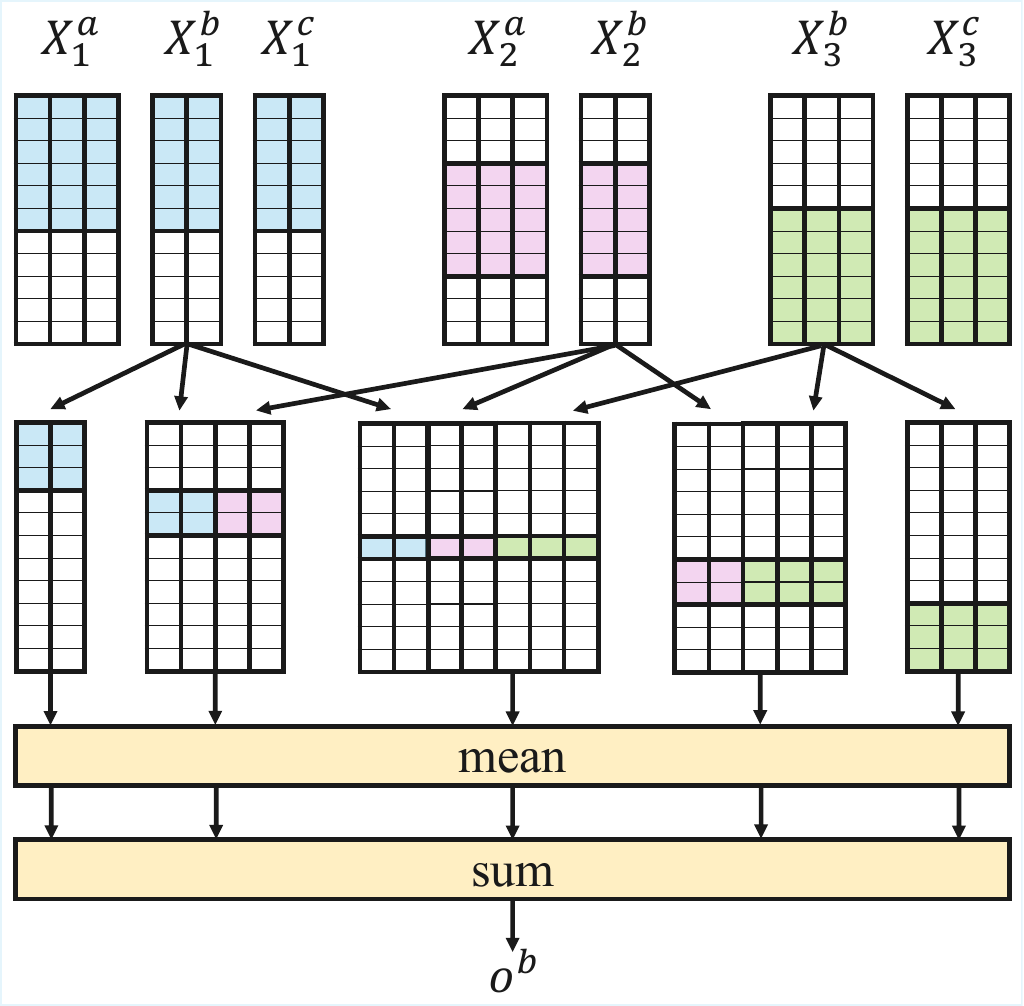}
    \caption{Example of computation of optimal cluster centroid, see \eqref{eq:optimal_masked_centroids}.}
    \label{fig:optimal_masked_centroids}
\end{figure}

Let $\mathcal{O} = \{O^1, ...,O^{K^\star}\}$ be the optimal clustering solution that would be obtained by applying a given distance-based clustering algorithm on the central dataset $\bar{X}$. $\mathcal{O}$ is a partition of $\bar{X}$ into $K^\star$ subsets, and $O^a \in \mathcal{O}$ is the $a$-th optimal cluster, also called true cluster. Given a set $S$, let $S^a = \bigl\{ x \in S \ \big| \ x \in O^a \bigr\}$. While the optimal clustering labels can be computed on $\bar{X}$, the corresponding optimal centroids should be computed by taking into account masking.
For each cluster $O^a \in \mathcal{O}$, the corresponding cluster centroid $o^a$ is defined by the entry-wise fraction
\begin{equation} \label{eq:optimal_masked_centroids}
    o^a = \ddfrac{ \sum_{i \in [n]} \sum_{x \in X_i^a} x }{ \sum_{m \in [d]} e_m \min \left( 1, \ \sum_{i \in [n]} \Omega_i[m,m] \cdot \big| X_i^a \big| \right) }.
\end{equation}
The above expression is the average over all the local data points belonging to the true cluster $O^a$, computed separately along each coordinate to consider only the entries that are not masked. A visual representation of this computation is provided in Figure \ref{fig:optimal_masked_centroids}.
We want to obtain a solution as close as possible to the partition $\mathcal{O}$ and the cluster centroids $\mathscr{o} = \{ o^1, ..., o^{K^\star} \}$, which are the best solution that can be attained.

\subsection{The challenge of partially overlapping feature spaces} \label{sec:no_uniform_scaling}

In this paper we propose two distributed algorithms to solve the above problem, both of which include an initial local clustering phase and a successive global aggregation. Ideally, we would like participants to correctly identify their local clusters, so that aggregation produces the correct global solution. The main problem with our setup is that, in general, masking shrinks the distances between pairs of data points without preserving their ratio. Some partial exceptions exist, such as the special cases mentioned below, where reducing from $d$ to $d_i$ dimensions can scale the relative distances between points almost uniformly.

\begin{itemize}
    \item Isotropy. In an isotropic distribution the covariance matrix is proportional to the identity matrix, which means that the variables are uncorrelated and the variance of the data is the same in all directions. If data are drawn from an isotropic distribution, then selecting a sufficiently large subset of coordinates should approximately preserve the ratio between pairwise distances. However, in several real-world applications the data exhibit various types of correlations and are not isotropic, e.g. \cite{de2019isotropy, arbia2008class}.
    \item Ad-hoc masking. If data lie near a lower-dimensional subspace or is sparse, and masking happens to capture the right subspace or sparsity pattern, then distances can be approximately retained. However, this is domain-specific and requires feature selection to choose the right subset of coordinates.
    \item Random projections. If masking is the result of a random projection, then Johnson-Lindenstrauss lemma \cite{johnson1984extensions} provides a formal guarantee on approximate pairwise distance preservation. The specific case of binary random projections is studied in \cite{li2022binary}, but this is different from masking as the projection matrix is not diagonal.
\end{itemize}

The above conditions are generally not verified in realistic settings, making it unreasonable to wish for the relative distances between data points to be scaled nearly uniformly. The fact that data are distributed over multiple participants further complicates things, as some clusters that could be easily identified in a centralized setting may not be recognizable by the individual participants.
To address these challenges, in this work we only require that the relative order between certain distances is preserved by masking, which is a much milder assumption with respect to assuming that all pairwise distances are scaled uniformly.

\subsection{Preliminary assumptions}

Below we introduce some assumptions characterizing data partitioning and masking, which will be recalled in the analysis of both proposed algorithms.
Depending on the specific combination of these and other assumptions, we will provide different theoretical arguments to support each algorithm. More specifically, we will identify sets of sufficient conditions under which each algorithm is expected to provide the optimal solution $\mathcal{O}, \mathscr{o}$.

Let $\mathcal{E} = \bigl\{ \overleftrightsmallarrow{ij} \ | \ (\Omega_i \wedge \Omega_j) \mathcal{X}^d \neq \{ 0^d \},\ i,j \in [n] \bigr\}$ be the set of edges that connect participants whose feature spaces overlap. Given any true cluster $O^a \in \mathcal{O}$, let $\mathcal{G}^a = ( \mathcal{V}^a, \mathcal{E}^a )$ be the undirected graph with vertices $\mathcal{V}^a = \bigl\{ i \in [n] \ | \ X_{i}^a \neq \varnothing \bigr\}$ and edges $\mathcal{E}^a = \bigl\{ \overleftrightsmallarrow{ij} \in \mathcal{E} \ | \ i,j \in \mathcal{V}^a \bigr\}$.

\begin{assumption}
\label{assumption:graph}
For any $a \in [K^\star]$:
\begin{enumerate}
    \item $\mathcal{G}^a$ is connected. \label{assumption_item:connected}
    \item $\displaystyle \bigvee_{i \in \mathcal{V}^a} \Omega_i = I$, the identity matrix. \label{assumption_item:identity_matrix}
    \item $\bigl\{ \bar{X}_i^a \ | \ i \in \mathcal{V}^a \bigr\}$ are identically distributed. \label{assumption_item:identically_distributed}
\end{enumerate}
\end{assumption}

In each graph $\mathcal{G}^a$, vertices are participants whose dataset contains data on the $a$-th true cluster, and edges connect participants whose feature spaces are not disjoint.
Item \ref{assumption_item:connected} of Assumption \ref{assumption:graph} says that the overlaps between the feature spaces of the participants allow for paths to be created that connect all participants who have data on the same true cluster.
Item \ref{assumption_item:identity_matrix} tells that, for each cluster, every feature is contained in at least one local dataset. Equivalently, for any $a \in [K^\star]$ the union of the feature spaces of the participants in $\mathcal{V}^a$ spans the whole $d$-dimensional space, so that information that might be relevant to identifying clusters is not totally lost.
Item \ref{assumption_item:identically_distributed} states that, leaving out masking, local data belonging to the same true cluster have the same distribution. Equivalently $\bar{X}_i^a \overset{d}{=} O^a$, $\forall i \in \mathcal{V}^a,\ \forall a \in [K^\star]$.
This facilitates collaboration between participants who have data on the same true cluster but on disjoint feature spaces.

\begin{assumption} \label{assumption:graph_fully_connected}
For any $a \in [K^\star]$:
\begin{enumerate}
    \item $\mathcal{G}^a$ is fully connected.
    \item $\displaystyle \bigvee_{i \in \mathcal{V}^a} \Omega_i = I$, the identity matrix.
\end{enumerate}
\end{assumption}

\begin{figure}[h]
    \centering
    \includegraphics[width=0.85\linewidth]{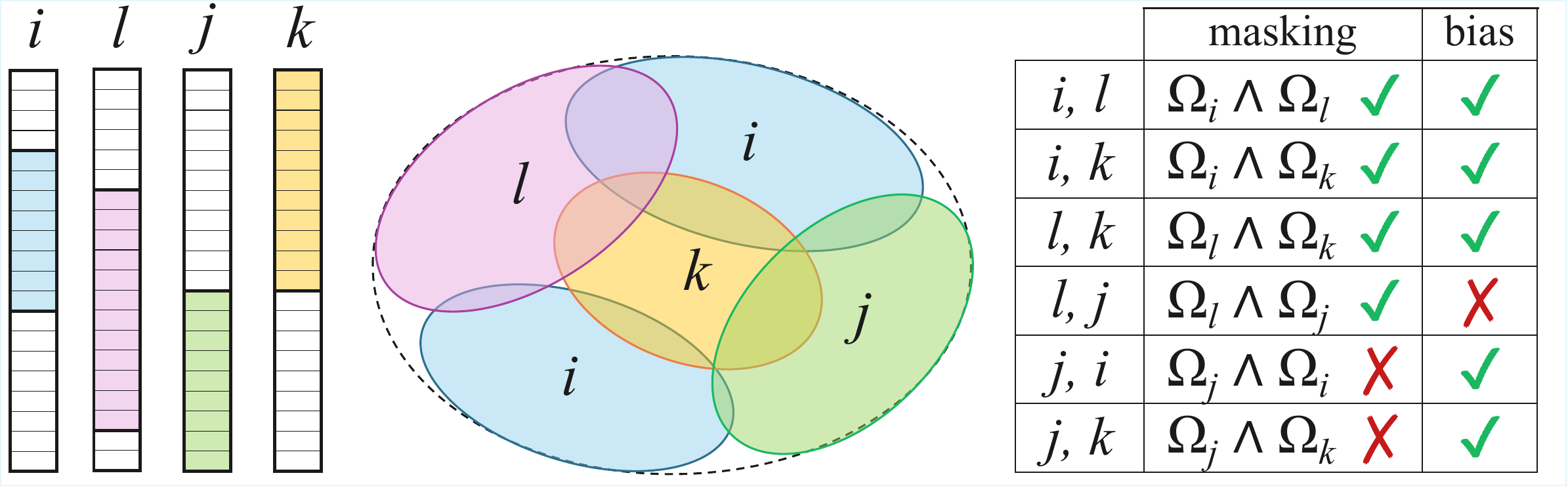}
    \caption{Situation in which the combination of masking and local bias due to heterogeneous data distributions prevents the merging of local clusters associated with the same true cluster. The colored entries in the column vectors on the left represent the feature spaces of the participants. Due to masking, the local solution of participant $j$ can only be compared with that of participant $l$, but due to bias, their local clusters are too different to be merged. The similarity between the local cluster of participant $j$ and the ones of participants $i$ and $k$ cannot be detected due to masking.}
    \label{fig:regional_merge_requires_all_intersections}
\end{figure}

Assumption \ref{assumption:graph_fully_connected} allows for heterogeneous data distributions, but in exchange requires that $\mathcal{G}^a$ be fully connected $\forall a \in [K^\star]$. This means that the feature spaces of any two participants who have data on the same true cluster are partially overlapping.
If the graphs were not fully-connected, participants whose local clusters should be merged may not have compatible feature spaces, as shown in the example in Figure \ref{fig:regional_merge_requires_all_intersections}.

\begin{remark}
The above assumptions should rather be considered sufficient conditions for distributed clustering to recover the optimal clustering solution $\mathcal{O}, \mathscr{o}$. In fact, if participants' feature spaces are incompatible, it is unreasonable to compare their local data without additional assumptions.
\end{remark}

Let $K_i$ be the number of clusters identified by participant $i$, and $K_i^\star$ be the correct number of clusters in the local dataset $X_i$ according to the target clustering solution $\mathcal{O}, \mathscr{o}$. Let $\mathcal{C}_i = \{ C_i^a \subseteq X_i,\ \forall a \in [K_i] \}$ such that $\bigcup_{a \in [K_i]} C_i^a = X_i$ be the local clustering solution to which participant $i$ converges.

\begin{assumption} \label{assumption:correct_local_overclustering}
     For any local cluster $C_i^a$ identified by any participant $i$, there is $b \in [K^\star]$ such that $C_i^a \subseteq X_{i}^b$.
\end{assumption}

Assumption \ref{assumption:correct_local_overclustering} states that each local cluster identified by the participants contains data points belonging to a single true cluster. As a consequence, merging the correct local clusters yields the correct global solution. We use this assumption to analyze the expected behavior of the proposed algorithms while allowing participants to perform local clustering using clustering algorithms of their choice. 
This assumption can always be satisfied by choosing $K_i$ sufficiently large, taking into account that $K_i \gg K_i^\star$ leads to excessive overclustering.

\subsection{Comparing distances in different feature spaces}

The proposed algorithms cluster data points based on their distance from each other.
We allow for a generic distance function $\phi(\cdot, \cdot): \mathcal{X}^d \times \mathcal{X}^d \to \mathbb{R}_+$, such as the standard Euclidean distance or the cosine distance.
When distances are computed in different feature subspaces, e.g. $(\Omega_i \wedge \Omega_j) \mathcal{X}^d$ and $(\Omega_j \wedge \Omega_k) \mathcal{X}^d$, they cannot be directly compared. To mitigate this problem, we rescale them by dividing by the maximum observed distance in the corresponding subspaces. For a subspace $(\Omega_j \wedge \Omega_k) \mathcal{X}^d$, the maximum observed distance on a set of points $S$ is
\begin{equation} \label{eq:d_max}
    d_{jk}^{\max}(S) = \phi \Bigl( (\Omega_j \wedge \Omega_k) \ewmax_{x \in S} x,\ (\Omega_j \wedge \Omega_k) \ewmin_{x' \in S} x \Bigr)
\end{equation}
where $\ewmax(\cdot)$ and $\ewmin(\cdot)$ return the entry-wise maximum and minimum, respectively. The distance between two generic data points $x \in \Omega_i \mathcal{X}^d$ and $x' \in \Omega_j \mathcal{X}^d$, rescaled according to the observations in $S$, is then
\begin{equation} \label{eq:d}
    d_{ij}^S(x, x') = \frac{ \phi \bigl( {(\Omega_i \wedge \Omega_j) x ,\ (\Omega_i \wedge \Omega_j) x'} \bigr) }{ d_{ij}^{\max}(S) }.
\end{equation}
If $\Omega_i \wedge \Omega_j$ is the zero matrix, then the feature spaces $\Omega_i \mathcal{X}^d$ and $\Omega_j \mathcal{X}^d$ are disjoint. In this case the distances $d_{ij}(\cdot, \cdot)$ return an invalid number and are ignored.
\section{Algorithm 1} \label{sec:federated_algorithm}

Algorithm \ref{alg:federated} is a federated clustering algorithm suited for participants with different but overlapping feature spaces. It aims at finding a given number $K$ of global centroids which are as close as possible to the optimal centroids $\mathscr{o}$. Since clustering is based on centroids, the algorithm should be applied to datasets where clusters are sufficiently described by centroids.
The algorithm is composed of an initialization phase, aimed at finding a starting set of global centroids, and of a federated phase, in which participants iteratively and collaboratively update these centroids.

\algorithmstyle{ruled}
\begin{algorithm}
\linespread{1.35}\selectfont
\DontPrintSemicolon
\SetNoFillComment
\SetArgSty{textnormal}
\SetCommentSty{mycommentfont}
parameters: global number of clusters $K$, stepsize $\alpha \in (0,1]$\par
\tcp{Initialization - Local clustering}
\For{each participant $i \in [n]$ in parallel}{
    labels, $\mathscr{c}_i \leftarrow$ localClustering($X_i$) \par
    }
\tcp{Initialization - Global aggregation at the server}
$\mathscr{c} \leftarrow \mathscr{c}_1 \cup \dots \cup \mathscr{c}_n$\par
\uIf{Assumption \ref{assumption:graph} is satisfied}{
    $G_c \leftarrow \text{methodA}(\mathscr{c}, K)$
    }
\uElseIf{Assumptions \ref{assumption:graph_fully_connected} and \ref{assumption:correct_local_overclustering} are satisfied}{
    $G_c \leftarrow \text{methodB}(\mathscr{c}, K)$
    }
\Else{
    warning(`case not supported, defaulting to methodB()')\par
    $G_c \leftarrow \text{methodB}(\mathscr{c}, K)$
    }
\tcp{Federated phase}
\While{not converged}{
    \For{each participant $i \in [n]$ in parallel}{
        $G_{ci} \leftarrow \bigl[ \ \Omega_i G_c[a], \ \forall a \in [K] \ | \ G_c[a] \text{ is compatible with } X_i \ \bigr]$\par
        labels, $\mathscr{c}_i \leftarrow$ localClustering($X_i$, init=$G_{ci}$) \par
        indices $\leftarrow$ sortHungarian$\left( \bigl\{ \Omega_i G_c[a], \ \forall a \in [K] \bigr\}, \mathscr{c}_i \right)$\par
        $\mathscr{c}_i' \leftarrow \mathscr{c}_i$\par
        \For{$a \in [K]$}{
            $c_i^a \leftarrow \mathscr{c}_i'[\text{indices}[a]] \textbf{ if }\text{indices}[a] \neq \nan \textbf{ else } \nan$
            }
        }
    $G_c \leftarrow (1-\alpha)G_c + \alpha \bigl[ \ \text{merge}( \{ c_i^a, \ \forall i \in [n] \ | \ c_i^a \neq \nan \} ) \ \forall a \in [K] \ \bigr]$\par    
    }
\Return $G_c$
\caption{Federated $K$-clustering of Masked Data}
\label{alg:federated}
\end{algorithm}

\subsection{Algorithm description}

\subsubsection{Initialization - local phase}

The initialization is a crucial step of most clustering algorithms and often determines the goodness of the resulting clustering solution. The simplest initialization, in which the initial $K$ centroids are randomly generated points, is not optimal in most cases. On the other hand, adopting more advanced initializations techniques such as the one of $K$-means\texttt{++} \cite{arthur2006k}, which chooses initial centroids sequentially, or \cite{yang2024greedy}, which proposes a greedy centroid initialization for federated $K$-means, is not straightforward in our setting.
Our proposed initialization is as follows. At the beginning of the algorithm, the central server broadcasts the maximum number of clusters $K$ and participants run an arbitrary clustering algorithm on their local dataset, individually and in parallel (lines 1-2).
Each participant $i$ can determine the optimal number of clusters $K_i \leq K$ for his local dataset using a standard method, such as maximizing the silhouette score \cite{rousseeuw1987silhouettes}. The resulting local centroids or medoids $\mathscr{c}_i = \bigl[ c_i^a, \ \forall a \in [K_i], \ K_i \leq K \bigr]$ are then sent to the server for aggregation.

\subsubsection{Initialization - global phase}

In this phase the server solves a meta-clustering problem, i.e. clustering the local clustering solutions. More specifically, the server aggregates the local centroids into $K$ global centroids, as also done in \cite{brandao2021efficient, holzer2023dynamically}.
Let $|c_i^a|$ denote the number of data points associated with centroid $c_i^a$, and define the function merge$(\cdot)$ as follows: 
\begin{equation} \label{eq:merge_centroids_function}
    \text{merge}(S) = \ddfrac{ \sum_{c_i^a \in S} c_i^a }{ \sum_{m \in [d]} e_m \min \left( 1, \ \sum_{c_i^a \in S} \Omega_i[m,m] \cdot |c_i^a| \right) }.
\end{equation}
This function computes the weighted average of the centroids in the input set $S$ and is similar to \eqref{eq:optimal_masked_centroids}.
Let $G$ be a list that encloses $K$ lists, where each list $G[g]$ contains the pairs (participant index, local cluster index) that identify the local centroids that compose the $g$-th global centroid, $\forall g \in [K]$. We formulate the meta-clustering problem as
\begin{equation} \label{eq:meta_clustering}
    G^\star = \argmin \sum_{g=1}^K \sum_{(i,a) \in G[g]} d_{ii}^c \Bigl( c_i^a,\ \text{merge} \bigl( \{ c_i^a, \ \forall (i,a) \in G[g] \} \bigr) \Bigr)^2.
\end{equation}
The above objective function seeks to minimize the variance of the local centroids that compose each global centroids. The distances are rescaled by the maximum observed distance computed on $\mathscr{c} = \bigl[ c_i^a, \ \forall i \in [n],\ \forall a \in [K_i] \bigr]$, which contains all the local centroids shared by the participants.
Depending on the characteristics of the clustering problem, we approximately solve \eqref{eq:meta_clustering} using two different methods.

\medskip
\textbf{Method A}

This method addresses the case where Assumption \ref{assumption:graph} is satisfied, i.e. $\bar{X}_i^a \overset{d}{=} O^a$, $\forall i \in \mathcal{V}^a,\ \forall a \in [K^\star]$. Since the local data distributions are identical at the cluster level, the optimal local centroids are unbiased estimates of the projected true centroids: $\mathbb{E}[o_i^a] = \Omega_i o^a$, $\forall i \in \mathcal{V}^a, \ \forall a \in [K^\star]$. This implies that the expected distance between optimal local centroids associated with the same true cluster is zero. 
Exploiting these facts, we propose  an iterative aggregation procedure where at each step we merge the two centroids with minimum distance.
If the local clustering solutions of the participants are sufficiently correct, the local centroids that are supposed to be merged should be at minimum distance, and the aggregation procedure should produce the desired result.
We can make the procedure more robust by noting that if the local clustering solution of a participant $i$ is sufficiently correct, it will contain exactly one centroid for every true cluster in the local dataset $X_i$. Consequently, it makes sense to solve \eqref{eq:meta_clustering} with a constraint: each global centroid cannot be made of multiple local centroids from the same participant.

\RemoveAlgoNumber
\algorithmstyle{ruled}
\begin{algorithm}
\SetAlgorithmName{methodA}
\DontPrintSemicolon
\SetNoFillComment
\SetArgSty{textnormal}
\SetCommentSty{mycommentfont}
\linespread{1.35}\selectfont
\SetKwInOut{Input}{Input}
\SetKwInOut{Output}{Output}
\Input{\hspace{0.7mm} Local centroids $\mathscr{c} = \bigl[ c_i^a, \ \forall i \in [n], \ \forall a \in [K_i] \bigr]$,\newline \hspace*{0.7mm} number of global centroids $1 \leq K \leq |\mathscr{c}|$.}\par
\Output{\hspace{0.7mm} Initial global centroids.}
$i \leftarrow$ index of any participant such that $K_i = K$\par
$G \leftarrow [ \ [(i,a)] \; \ \forall a \in [K] \ ]$\par
$G_c \leftarrow [ \ (c_i^a, \Omega_i) \; \ \forall a \in [K] \ ]$\par
$L \leftarrow [ \ (j,b) \; \ \forall j \in [n], \ j \neq i, \ \forall b \in [K_j] \ ]$\par
\While{$|L| > 1$}{
    \For(\tcp*[f]{naive implementation}){$l \in [|L|]$}{
        $(j,b) \leftarrow L[l]$\par
        \For{$g \in [K]$}{
            \uIf{$i \neq j \ \forall(i,a) \in G[g]$}{
                $(c_i^a, \Omega_i) \leftarrow G_c[g]$\par
                $D_{lg} \leftarrow d_{ij}(c_i^a,\ c_j^b)$\par
                }
            \Else{
                $D_{lg} \leftarrow \infty$
                }
            }
        }
    $(l,g) \leftarrow \argmin \left( \{ D_{lg}, \ \forall l \in [|L|], \ \forall g \in [K] \} \right)$\par
    $G[g].\text{append}(L[l])$\par
    $(j,b) \leftarrow L.\text{pop}(l)$\par
    $(c_i^a, \Omega_i) \leftarrow G[g]$\par
    $G_c[g] \leftarrow \left( \text{merge}( \{ c_i^a, c_j^b \} ),\ \Omega_i \vee \Omega_j \right)$\par    
    }
\Return the centroids in $G_c$
\caption*{Approximate meta-clustering under Assumption \ref{assumption:graph}}
\label{alg:methodA_meta_clustering}
\end{algorithm}
\RevertAlgoNumber

We now describe the merging procedure in detail. Merging must satisfy two conditions: at the end of the process we want to obtain $K$ global centroids, and each global centroid must be a combination of local centroids from different participants. For simplicity, we assume that there is a participant $i$ having exactly $K_i = K$ clusters, and we initialize $G$ using his local centroids (line 2). If this is not the case, $G$ should be initialized using $K$ local centroids that with high probability should not be merged.
The list $G_c$ is an alternative representation of the current global centroids in $G$, and contains the actual vectors and the associated projection matrices (line 3). The list $L$ contains the pairs (participant index, local cluster index) that are not in $G$ (line 4). At each merging iteration, we compute the rescaled distance between each global centroid in $G$ and each local centroid in $L$ in the intersection between the associated feature spaces (line 11). If the associated feature spaces are disjoint, the distance is not computed. If the global centroid already contains a local centroid from the same participant, the distance is set to infinity to prevent merging (line 13). We then merge the global centroid and the local centroid with minimum distance, updating the various lists (lines 14-18). 
When the list $L$ is empty, the merging procedure is completed and $G_c$ contains the global centroids.

\medskip
\textbf{Method B}

\RemoveAlgoNumber
\algorithmstyle{ruled}
\begin{algorithm}
\SetAlgorithmName{methodB}
\DontPrintSemicolon
\SetNoFillComment
\SetArgSty{textnormal}
\SetCommentSty{mycommentfont}
\linespread{1.35}\selectfont
\SetKwInOut{Input}{Input}
\SetKwInOut{Output}{Output}
\Input{\hspace{0.7mm} Local centroids $\mathscr{c} = \bigl[ c_i^a, \ \forall i \in [n], \ \forall a \in [K_i] \bigr]$,\newline \hspace*{0.7mm} number of global centroids $1 \leq K \leq |c|$.}\par
\Output{\hspace{0.7mm} Initial global centroids.}
%
\tcp{Find the two centroids at maximum distance}
distanceMatrix $\leftarrow$ $0^{|c| \times |c|}$, the zero matrix\par
\For{$z_1 \in [|\mathscr{c}|]$}{
    \For{$z_2 \in [|\mathscr{c}|]$, $z_2 > z_1$}{
        $c_i^a \leftarrow \mathscr{c}[z_1]$, $c_j^b \leftarrow \mathscr{c}[z_2]$\par
        distanceMatrix$[z_1, z_2] \leftarrow d_{ij}(c_i^a,\ c_j^b)$\par
        distanceMatrix$[z_2, z_1] \leftarrow \text{distanceMatrix}[z_1, z_2]$
        }
    }
$\displaystyle (z_1, z_2) \leftarrow \argmax_{z_1, z_2} \text{ distanceMatrix}[z_1, z_2]$\par
$Z \leftarrow \{ z_1, z_2 \}$\par
\tcp{Iteratively add the centroid at max-min distance}
\For{$k \in [3,K]$}{
    minDistances $\leftarrow 0^{|c|}$, the zero vector\par
    \For{$z \in [|\mathscr{c}|] \backslash Z$}{
        minDistances$\displaystyle [z] \leftarrow \min_{i \in Z} \text{ distanceMatrix}[z, i]$
        }
    $\displaystyle z_k \leftarrow \argmax_z \text{ minDistances}[z]$\par
    $Z \leftarrow Z \cup z$
    }
\tcp{Cluster local centroids using $K$-means}
$G_c \leftarrow [ \ (c_i^a = \mathscr{c}[z],\ \Omega_i) \; \ \forall z \in Z \ ]$\par
\While{not converged}{
    clusters $\leftarrow [ \ \{ \} \; \ \forall k \in [K] \ ]$\par
    masks $\leftarrow [ \ 0^{d \times d} \; \ \forall k \in [K] \ ]$\par
    \For{$c_j^b \in \mathscr{c}$}{
        distances $\leftarrow 0^{K}$, the zero vector\par
        \For{$k \in [K]$}{
            $(c_i^a, \Omega_i) \leftarrow G_c[k]$\par
            distances$[k] \leftarrow d_{ij}^c (c_i^a, c_j^b)$\par
            }
        $k \leftarrow \argmin$ distances \par
        $\text{clusters}[k] \leftarrow \text{clusters}[k] \cup c_j^b$\par
        $\text{masks}[k] \leftarrow \text{masks}[k] \vee \Omega_j$\par
        }
    $G_c \leftarrow \left[ \ \left( \text{merge}( \text{clusters}[k] ),\ \text{masks}[k] \right) \ \forall k \in [K] \ \right]$\par  
    }
\Return the centroids in $G_c$
\caption*{Approximate meta-clustering under Assumptions \ref{assumption:graph_fully_connected} and \ref{assumption:correct_local_overclustering}}
\label{alg:second_method_meta_clustering}
\end{algorithm}
\RevertAlgoNumber

\begin{figure}[htb]
    \centering
    \includegraphics[width=0.5\linewidth]{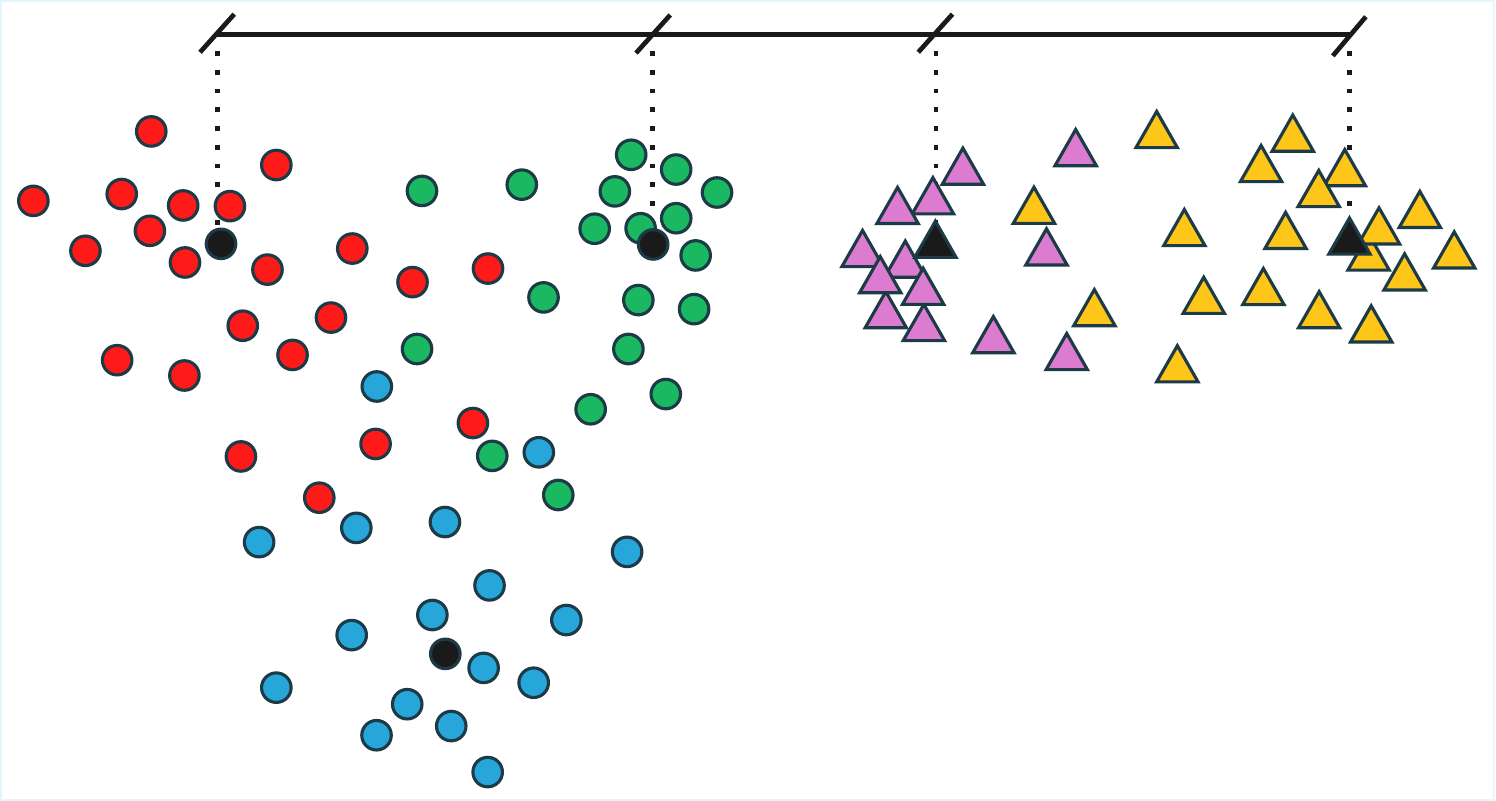}
    \caption{In the figure, data points with different shapes belong to different true global clusters. Different colors indicate different local clusters, and the corresponding centroids are shown in black. Both local clusters and true global clusters are well described by their centroids. In this example, aggregating the two centroids at minimum distance (green dots and purple triangles) results in wrong clustering. This is due to the fact that local data belonging to the same true cluster have very different distributions.}
    \label{fig:centroids_need_uniform}
\end{figure}

The second method addresses the case where Assumptions \ref{assumption:graph_fully_connected} and \ref{assumption:correct_local_overclustering} are satisfied. Since local versions of the same cluster may have very different data distributions, iteratively aggregating the two local centroids at minimum distance is not appropriate in this case, as shown in Figure \ref{fig:centroids_need_uniform}.

In this case, we propose to cluster local centroids using a modified version of $K$-means. First we solve the following maximum minimum diversity problem: find the $K$ local centroids that are most distant from each others. We approximately solve this NP-hard problem greedily, iteratively selecting the local centroid that is most distant from the previously selected ones. Then we apply $K$-means, initialized using the local centroids at maximum minimum distance, whose indices are contained in the set $Z$. To adapt $K$-means to our needs, we compute distances between local centroids using \eqref{eq:d}, and global centroids using \eqref{eq:merge_centroids_function}. 
Overall, Method B can be seen as a deterministic and appropriately adapted version of $K$-means\texttt{++} \cite{arthur2006k}. The output of Method B is the set $G_c$, which contains the initial global centroids.

\subsubsection{Federated phase}

For well-conditioned clustering problems in which clusters are sufficiently separated, $G_c$ may already be a good global solution. To further improve it, we propose a federated iteration that can be repeated until a given convergence criterion is satisfied. At each federated iteration participants update their local clustering solution, individually and in parallel, performing the following steps.
\begin{enumerate}
    \setlength\itemsep{-1mm}
    \item Each participant $i$ selects the current global centroids in $G_c$ that are compatible with his local dataset, obtaining $G_{ci}$ (line 14). Assuming that each local cluster contains at least a certain number of points, compatibility can be assessed by assigning each point in the local dataset $X_i$ to the closest global centroid, and selecting only the global centroids associated with a sufficient number of points.
    \item Participants run again an arbitrary clustering algorithm on their local dataset for a small number of iterations. The chosen clustering algorithm must allow setting the initial centroids, so that each participant $i$ can initialize the centroids using $G_{ci}$ (line 15). If the chosen algorithm requires medoids instead of centroids, participants can set their local data points closest to the selected centroids as initial medoids.
    \item Each participant $i$ uses the Hungarian method \cite{kuhn1955hungarian} to find the minimum cost matching between the updated local centroids $\mathscr{c}_i$ and the global centroids $G_c$, where the cost of matching two centroids is simply their distance. The minimum cost solution is used to align the $K_i$ local centroids with the $K$ global ones, obtaining $K$ aligned local centroids. If the $a$-th global centroid is not matched with any local centroid, the $a$-th aligned local centroid is set equal to an invalid value such as \nan, standing for \doubleapex{not a number} (lines 16-18).
    \item Finally, the aligned local centroids are sent to the server for aggregation, where only valid centroids are merged. The update rule computes a convex combination of the old centroids and the new ones, where the stepsize $\alpha \in (0, 1]$ is a tunable parameter (line 19).
\end{enumerate}

\subsection{Algorithm analysis}

To analyze the proposed algorithm, let us pretend that the local datasets result from the procedure described in Section \ref{sec:problem_formulation}. We want to find some conditions under which we can reasonably expect Algorithm \ref{alg:federated} to converge to the optimal clustering solution $\mathcal{O}, \mathscr{o}$.

\subsubsection{Analysis of the initialization - local phase} \label{sec:alg_1_analysis_local_clustering}

In this phase participants perform clustering on their local datasets using clustering algorithms of their choice.
Since we allow participants to choose arbitrary clustering algorithms, we cannot provide exact guarantees on the correctness of the local clustering solutions to which the participants converge. Depending on the characteristics of the problem we consider two scenarios, corresponding to Assumptions \ref{assumption:graph} and \ref{assumption:graph_fully_connected}.

To reasonably expect that participants are able to correctly identify their local clusters, we need the combination of data partitioning and masking to leave the true clusters $\mathcal{O}$ sufficiently separated.
Under Assumption \ref{assumption:graph}, data partitioning satisfies $\bar{X}_i^a \overset{d}{=} O^a$, $\forall i \in \mathcal{V}^a,\ \forall a \in [K^\star]$. Consequently, the true local clusters $X_i^1, \dots, X_i^{K^\star}$ of each participant $i$ should be almost as compact and separated as in the centralized dataset $\bar{X}$. This is especially true in the high-data regime, where each non-empty true local cluster contains enough data points.
As for masking, feature removal typically shrinks the distances between data points without preserving their proportions, as previously discussed in Section \ref{sec:no_uniform_scaling}. For this reason, we characterize masking using the following assumption, where $\convexhull(S)$ is the convex hull of a generic dataset $S$.

\begin{assumption} [Local conservation of the order between distances] \label{assumption:LCOD}
    $\forall a,b \in [K^\star],\ b \neq a$ and $\forall i \in \mathcal{V}^a \cap \mathcal{V}^b$ it holds:
    \begin{equation} \label{eq:LCOD}
        \phi(x_1, x_2) < \phi(x_1, x_3) \Rightarrow \phi (\Omega_i x_1, \Omega_i x_2) < \phi (\Omega_i x_1, \Omega_i x_3)
    \end{equation}
    $$\forall x_1 \in \bar{X}^a,\ x_2 \in \convexhull \bigl( \bar{X}^a \bigr),\ x_3 \in \convexhull \bigl( \bar{X}^b \bigr).$$
\end{assumption}

\begin{figure}[htb]
    \centering
    \includegraphics[width=0.5\linewidth]{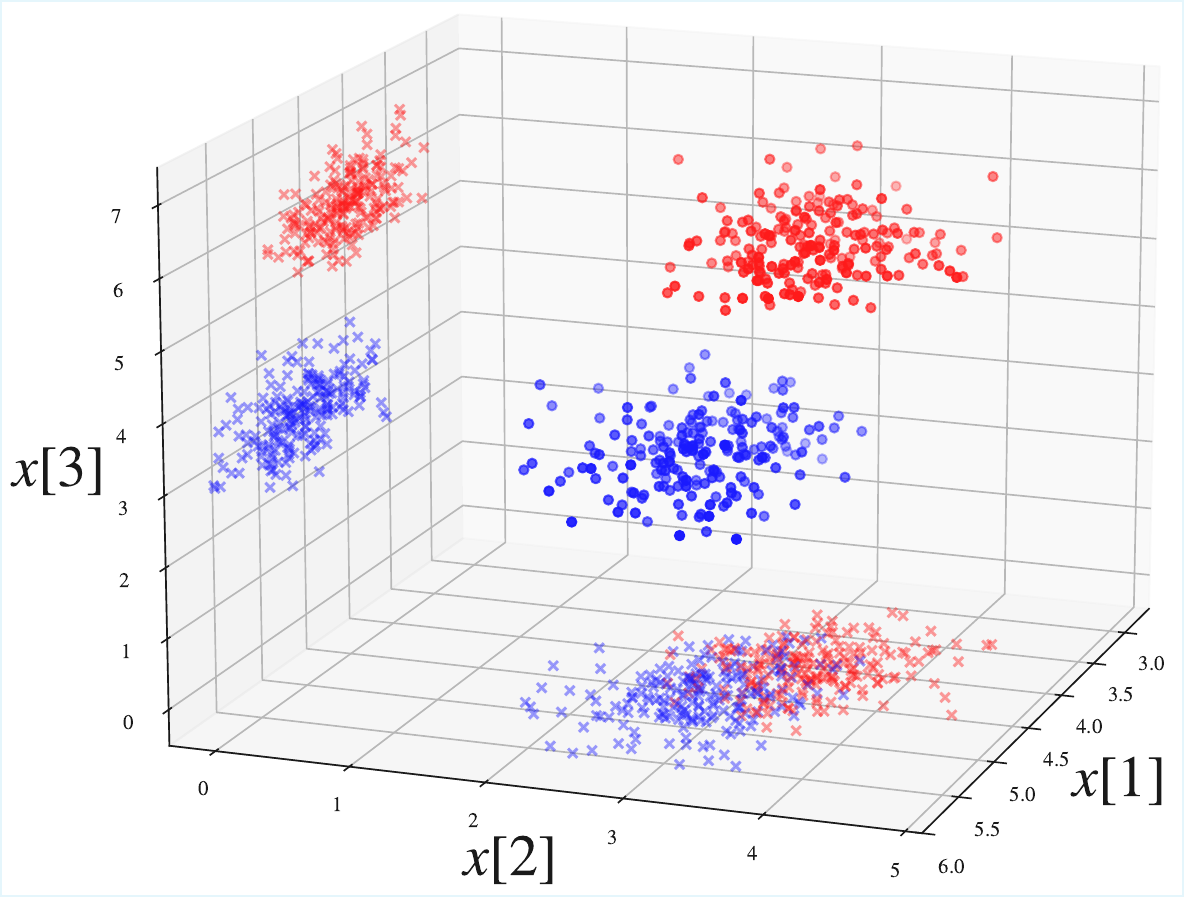}
    \caption{Intuition behind Assumption \ref{assumption:LCOD}. Masking the second coordinate of the data points leaves clusters well separated, while masking the third coordinate makes them seem like a single cluster in the lower-dimensional subspace. The implication \eqref{eq:LCOD} ensures that the projection matrices do not mix together data points belonging to different clusters.}
    \label{fig:good_and_bad_masking}
\end{figure}

Assumption \ref{assumption:LCOD} states that the axis-aligned projections induced by the projection matrices of the individual participants preserve the relative order of the distances between certain triplets of points. This indirectly guarantees that the information loss due to masking does not cause mixing of different clusters, as suggested by Figure \ref{fig:good_and_bad_masking}. The requirement is not too restrictive, since it concerns only the relative order between the distances rather than their value, and the implication need only hold for pairs of points that lie in the same cluster and that in the original space are closer to each other than to the third point.

To see the practical usefulness of Assumption \ref{assumption:LCOD}, imagine a generic participant $i$ performing local clustering on the non-masked dataset $\bar{X}_i$ using $K$-means or a similar distance-based clustering algorithm. $K$-means algorithms alternate between assignment steps, in which each point is assigned to the cluster with the nearest centroid, and update steps, in which the centroid of each cluster is recomputed. 
Consider a local data point $x \in \bar{X}_i$ belonging to any cluster $O^a$. By Assumption \ref{assumption:LCOD}, for any two points $c' \in \convexhull \bigl( \bar{X}^a \bigr)$, $c'' \in \convexhull \bigl( \bar{X}^b \bigr)$, $b \neq a$, if $\phi(x, c') < \phi(x, c'')$ then $\phi (\Omega_i x, \Omega_i c') < \phi (\Omega_i x, \Omega_i c'')$. Choose $c'$ and $c''$ to be two centroids: this means that if point $x$ is correctly assigned to cluster $O^a$ in the fully observable $d$-dimensional feature space, then it is correctly assigned also in the masked subspace $\Omega_i \mathcal{X}^d$.

Since participants can freely choose the algorithm to use for local clustering, they can opt for computationally inexpensive algorithms that can be repeated multiple times with different initializations. Repeating local clustering multiple times reduces the likelihood of errors and makes convergence to the optimal local solution plausible. In particular, if participants perform local clustering using the same clustering algorithm used to compute $O$, under Assumptions \ref{assumption:graph} and \ref{assumption:LCOD} they should converge to the optimal or to a near-optimal clustering solution.

Under Assumption \ref{assumption:graph_fully_connected}, data belonging to the same true cluster may have different distributions for different participants, and it may be difficult to recognize true local clusters. Furthermore, participants might split a single true cluster into multiple biased sub-clusters. In this case, providing guarantees that are valid for all clustering algorithms that participants may use is difficult. For the purposes of subsequent analysis, we resort to Assumption \ref{assumption:correct_local_overclustering}, which ensures that all clusters identified by the participants are pure, i.e. they contain data points from a single true cluster.

\subsubsection{Analysis of the initialization - global phase}

In this phase the server aggregates the local centroids based on their distances from each other.
We need to ensure that although distances between local centroids are computed in different feature subspaces, e.g. $(\Omega_i \wedge \Omega_j) \mathcal{X}^d$ and $(\Omega_i \wedge \Omega_k) \mathcal{X}^d$, they can still be meaningfully compared.
To do so, we introduce the function $\mho(\cdot): \Omega \mathcal{X}^d \mapsto \mathcal{X}^d$, which simulates what would happen if data were not masked. More specifically, if $x=(\psi \circ \text{mask})(S)$ is the result of masking a set of points $S \subseteq \bar{X}$ and applying an arbitrary function $\psi(\cdot)$ to them, $\mho(x) = \psi(S)$, without masking the input points.
\begin{equation} \label{eq:mho}
\begin{aligned}
    x &= \psi \bigl( \{ \ \Omega x' \; \ \forall x' \in S \subseteq \bar{X} \ | \ \Omega \in \{0, 1\}^{d \times d} \text{ is diagonal} \ \} \bigr) \\
    \mho(x) &= \psi(S).
\end{aligned}
\end{equation}

\medskip
\textbf{Analysis of Method A}

Method A relies on Assumption \ref{assumption:graph} to iteratively merge the closest local centroids. We want to ensure that local centroids associated with the same true cluster that in the absence of masking are closer to each other than to other local centroids remain closer even when data are masked. To this aim, we introduce Assumption \ref{assumption:for_methodA}, paraphrased below. For any $a\neq b \in [K^\star]$, consider any two participants $i,j$ having data on the true cluster $O^a$ and any participant $k$ having data on the true cluster $O^b$, such that the feature spaces of $j$ and $k$ overlap with that of $i$. This formulation simply considers all possible combinations of $a,b,i,j,k$ that can occur. Suppose that, if centroids were calculated using unmasked data, two local centroids of participants $i$ and $j$ associated with the same true cluster $O^a$ would be closer to each other than to a local centroid of participant $k$ associated with the true cluster $O^b$. The assumption states that data masking preserves this proximity relationship between local centroids.

\begin{assumption} \label{assumption:for_methodA}
    Let $\mathscr{c}_i$ be the set of local centroids shared by participant $i$. $\forall a,b \in [K^\star],\ b \neq a$, and $\forall i \in \mathcal{V}^a$, $\forall j \in \mathcal{V}^a$, $\forall k \in \mathcal{V}^b$ such that $j,k \in \mathcal{N}(i) = \{ v \in [n] \ | \ \overleftrightsmallarrow{iv} \in \mathcal{E} \}$ it holds:
    \begin{equation} \label{eq:for_methodA}
        \phi \bigl( \mho(x_1), \mho(x_2) \bigr) < \phi \bigl( \mho(x_1), \mho(x_3) \bigr) \Rightarrow d_{ij}(x_1, x_2) < d_{ik}(x_1, x_3)
    \end{equation}
    \begin{equation*}
        \begin{aligned}
            &\forall x_1 \in \bigl\{ x \in \mathscr{c}_i \ | \ \mho(x) \in \convexhull \bigl( \bar{X}^a \bigr) \bigr\}, \\
            &\forall x_2 \in \bigl\{ x \in \mathscr{c}_j \ | \ \mho(x) \in \convexhull \bigl( \bar{X}^a \bigr) \bigr\}, \\
            &\forall x_3 \in \bigl\{ x \in \mathscr{c}_k \ | \ \mho(x) \in \convexhull \bigl( \bar{X}^b \bigr) \bigr\}.
        \end{aligned}
    \end{equation*}
\end{assumption}

We comment on the expected performance of Method A under Assumptions \ref{assumption:graph}, \ref{assumption:LCOD} and \ref{assumption:for_methodA}.
Based on Assumption \ref{assumption:graph}, the overlaps between the feature spaces of the participants do not prevent the computation of necessary distances between local centroids, allowing all data points in the local datasets to be exploited.
Since the local datasets are identically distributed at the cluster level, the information loss due to masking does not introduce bias, and the available information should be sufficient to recover the correct cluster memberships $O$.
As previously argued, since the local versions $\bar{X}_i^a, \dots, \bar{X}_n^a$ of the same true cluster have identical distributions, the local centroids from different participants associated with the same true cluster should be very similar. 
Based on Assumption \ref{assumption:for_methodA}, we can compare the distances between the local centroids of different pairs of participants even if they are computed in different feature spaces. Since the local centroids that are supposed to be merged should be at minimum distance, Method A should produce the desired result.

\medskip
\textbf{Analysis of Method B}

Method B relies on Assumptions \ref{assumption:graph_fully_connected} and \ref{assumption:correct_local_overclustering} to aggregate local centroids using an adapted version of $K$-means. In the initialization, Method B identifies the set of local centroids at maximum minimum distance from each other. To compute distances between local centroids, we resort to Assumption \ref{assumption:for_methodA}. In the assignment steps of the adapted $K$-means, Method B computes distances between local centroids and aggregated local centroids. This leads us to introduce Assumption \ref{assumption:for_methodB}.

\begin{assumption} \label{assumption:for_methodB}
    Let $\mathscr{c}_i$ be the set of local centroids shared by participant $i$. $\forall a,b \in [K^\star],\ b \neq a$, and $\forall i \in \mathcal{V}^a$ it holds:
    \begin{equation} \label{eq:for_methodB}
        \phi \bigl( \mho(x_1), \mho(x_2) \bigr) < \phi \bigl( \mho(x_1), \mho(x_3) \bigr) \Rightarrow \phi (x_1, \Omega_i x_2) < \phi (x_1, \Omega_i x_3)
    \end{equation}
    \begin{equation*}
        \begin{aligned}
            &\forall x_1 \in \bigl\{ x \in \mathscr{c}_i \ | \ \mho(x) \in \convexhull \bigl( \bar{X}^a \bigr) \bigr\}, \\
            &\forall x_2 \in \bigl\{ x \ | \ \mho(x) \in \convexhull \bigl( \bar{X}^a \bigr) \bigr\}, \\
            &\forall x_3 \in \bigl\{ x \ | \ \mho(x) \in \convexhull \bigl( \bar{X}^b \bigr) \bigr\}.
        \end{aligned}
    \end{equation*}
\end{assumption}

We now comment on the expected performance of Method B under Assumptions \ref{assumption:graph_fully_connected}, \ref{assumption:correct_local_overclustering}, \ref{assumption:for_methodA} and \ref{assumption:for_methodB}.
Based on Assumption \ref{assumption:graph_fully_connected}, the feature spaces of the participants allow computing distances between any two local centroids, ensuring that all data points are used.
Although clusters may have different local data distributions, Assumption \ref{assumption:correct_local_overclustering} ensures that aggregating the correct local centroids provides the correct solution. Under Assumption \ref{assumption:for_methodB}, data masking does not significantly affect the assignment step of the modified $K$-means, and therefore Method B should return the correct aggregation.

\subsubsection{Analysis of the federated phase}

At each federated iteration, participants perform a compatibility check that favors they contribute only to global clusters for which they have data points.
Next, participant update their local clustering solution, and the analysis is similar to the one in Section \ref{sec:alg_1_analysis_local_clustering}. In particular, if participants perform a limited number of local clustering updates, their local solution should not deviate significantly from the current global solution.
Even in the event of significant changes, the alignment based on the Hungarian method maximizes the chances of correct global aggregation.
All the previous steps rely on calculating distances in the same feature space, and Assumption \ref{assumption:LCOD} ensures that these distances are meaningful and can be compared.
Additionally, the federated phase allows for the correction of any errors in the initialization phase.

\subsubsection{Limitations of Algorithm \ref{alg:federated}}

The main limitation of Algorithm \ref{alg:federated} is that it implicitly assumes that clusters are well described by their centroids. This is true when clusters are sufficiently separated or resemble spherical blobs, but it not true in some special cases, e.g. when clusters are concentric circumferences. 
In the latter case, even with a single centralized dataset and fully observable data, most distance-based clustering algorithms fail and other approaches, such as spectral clustering or DBSCAN \cite{ester1996density}, are more suitable.
To address scenarios in which local clusters have complex shapes, in the next section we propose a second algorithm based on a different approach.

\subsection{Computational complexity}

We estimate the computational cost of Algorithm \ref{alg:federated} in terms of time complexity. For simplicity, suppose that all participants always have $K_i = K$ clusters.

We start by computing the cost for each participant. In the initialization phase, the cost for participant $i$ is simply the cost of running a clustering algorithm on the local dataset $X_i$, containing $N$ data points. The associated computational cost depends on the chosen algorithm: in the case of $K$-means, each iteration requires the computation of $KN$ distances with complexity $\Theta(d_i)$ each, resulting in a time complexity per iteration $\Theta(d_i K N)$ \cite{schutze2008introduction}.
At each federated iteration, participants perform clustering on their local dataset for a relatively small number of iterations, starting from a subset of the current global centroids (line 25 of the algorithm). If the selection of the compatible global centroids involves computing distances between each data point in $X_i$ and each global centroid, this adds a complexity $\Theta(d_i KN)$. The updated local centroids are then aligned with the global ones by computing $K^2$ distances and using the Hungarian algorithm, whose complexity is $O(K^3)$ \cite{edmonds1972theoretical}. Consider a generic participant $i$ that uses $K$-means to perform local clustering, running $I_0$ $K$-means iterations in the initialization phase and $I_F \leq I_0$ $K$-means iterations at each federated iteration: with $F$ federated iterations, the overall complexity of Algorithm \ref{alg:federated} for participant $i$ is $\Theta(d_i KNI_0) + F \Theta \bigl( d_i KN(1 + I_F) + K^2 d_i + K^3 \bigr)$.

We now estimate the cost for the central server. 
When using Method A, at each iteration in the while loop, the server computes the distances between the $K$ global centroids in $G_c$ and the local centroids in $L$, looks for the smallest distance and merges two centroids. Since at the first iteration $|L| = K(n-1)$, the server computes at most
$$ K^2(n-1) + \sum_{i=1}^{K(n-1)-1} i = \Theta(K^2 n^2)$$
distances, performs $O(K^2 n^2)$ scalar comparisons to find the smallest distances, and calls the function merge$(\cdot)$ $K(n-1)-1$ times, resulting in a complexity $\Theta(d K^2 n^2)$.
When using Method B, the server calculates the distances between all pairs of local centroids, with a cost of $\Theta(d K^2 n^2)$. Then, the server runs a modified version of $K$-means to cluster the $Kn$ local centroids: each iteration requires the computation of $K^2 n$ distances with complexity $\Theta(d)$ each, resulting in a time complexity per iteration $\Theta(d K^2 n)$.
At each federated iteration, the central server calls the function merge$(\cdot)$ $K$ times, merging $n$ centroids at the time. When using Method A, the overall cost for the server is therefore $\Theta(d K^2 n^2 + FdKn)$.
\section{Algorithm 2} \label{sec:one_shot_algorithm}

Algorithm \ref{alg:one_shot} consists of a local phase, in which participants perform clustering on their private dataset and fit a probability distribution on each cluster, and a global aggregation phase, in which local clusters are merged. Notably, the algorithm is one-shot, meaning it requires a single communication round between participants and the central server, significantly reducing communication overhead compared to iterative approaches.

\subsection{Merging sets of clusters} \label{sec:attractive_force}

Below we introduce the procedure for merging local clusters used in the global aggregation phase of Algorithm \ref{alg:one_shot}.
We propose to merge clusters based on a novel physics-inspired mechanism which resembles gravitational or magnetic force. The similarity between data points is modeled as an attractive force: any two data points attract each others with a force that is inversely proportional to a power of their distance. We let the power exponent $w > 1$ be a tunable parameter, recovering inverse-square laws for $w=2$. Larger values of $w$ assign a greater importance to neighboring data points, and minimize the influence of distant data points.
Let $\{ P_i^a, \ a \in [K_i] \}$ be the local clusters of participant $i$, and $P$ be the union of the local clusters from all participants.
The unnormalized attractive force $f'(\cdot, \cdot)$ and the normalized attractive force $f(\cdot, \cdot)$ between a cluster $P_i^a$ coming from participant $i$ and a cluster $P_j^b$ from participant $j$ are
\begin{equation} \label{eq:attractive_force_two_clusters}
    f'(P_i^a, P_j^b)= \sum_{x \in P_i^a} \sum_{x' \in P_j^b} \frac{1}{ d_{ij}^P (x, x')^w },
    \qquad
    f(P_i^a, P_j^b) = \frac{ f'(P_i^a, P_j^b) }{|P_i^a| |P_j^b|}.
\end{equation}
Let the list $G[r] = [(i,a), (j,b), \dots]$ identify the set of clusters $\{ P_i^a,\ \forall (i,a) \in G[r] \}$. We use this notation to be consistent with that used in the \ref{alg:one_shot} algorithm.
The normalized attractive force between two sets of clusters defined by the lists $G[r]$ and $G[s]$ can be efficiently computed as
\begin{equation} \label{eq:attractive_force_sets_of_clusters}
    f (G[r], G[s]) = \ddfrac{ \sum_{(i,a) \in G[r]} \sum_{(j,b) \in G[s]} f'(P_i^a, P_j^b) } { \sum_{(i,a) \in G[s]} |P_i^a| \sum_{(j,b) \in G[s]} |P_j^b| } .
\end{equation}
The proposed aggregation procedure consists in iteratively merging the two sets of clusters with the highest mutual force of attraction \eqref{eq:attractive_force_sets_of_clusters}.
It begins by considering each cluster as a set of clusters containing only one element, and iteratively merges sets of clusters in a hierarchical bottom-up fashion.
The procedure terminates when the number of sets of clusters reaches the optimal number of clusters, which can be given or estimated via standard techniques, e.g. using the silhouette score \cite{rousseeuw1987silhouettes} to determine whether merging two sets of clusters is beneficial.

\begin{figure}[h]
    \centering
    \begin{tabular}{@{}cc@{}}
    \toprule
    distance between $G[r], G[s]$ & \includegraphics[width=3mm, height=3mm]{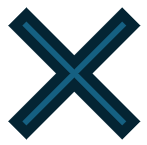} is merged with \\ \midrule
    $ \displaystyle \min_{(i,a) \in G[r]} \min_{(j,b) \in G[s]} \min_{x \in P_i^a} \min_{x' \in P_j^b} d_{ij}(x, x') $ & \includegraphics[width=3mm, height=3mm]{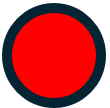}  \\
    $ \displaystyle \min_{(i,a) \in G[r]} \min_{(j,b) \in G[s]} \frac{1}{|P_i^a| |P_j^b|} \sum_{x \in P_i^a} \sum_{x' \in P_j^b} d_{ij}(x, x') $ & \includegraphics[width=9mm, height=4.8mm]{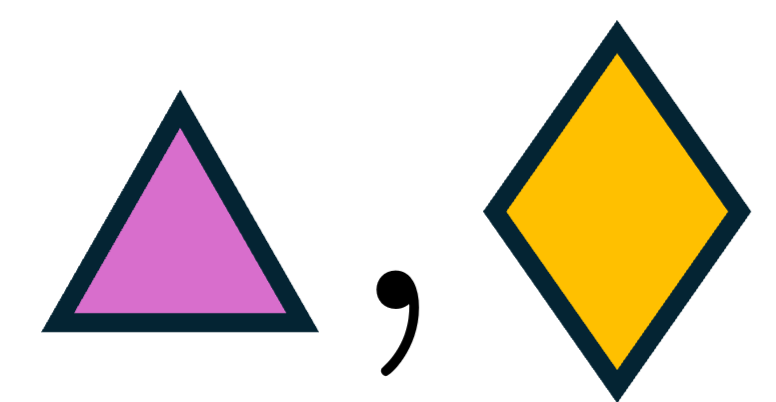}  \\
    $ \displaystyle \frac{1}{\text{\#tot}}  \sum_{(i,a) \in G[r]} \sum_{(j,b) \in G[s]} \sum_{x \in P_i^a} \sum_{x' \in P_j^b} d_{ij}(x, x') $ & \includegraphics[width=9mm, height=4.8mm]{fig/purple_triangle_orange_diamond.png} \\
    $ \displaystyle - f(G[r], G[s]) $ & \includegraphics[width=9mm, height=4mm]{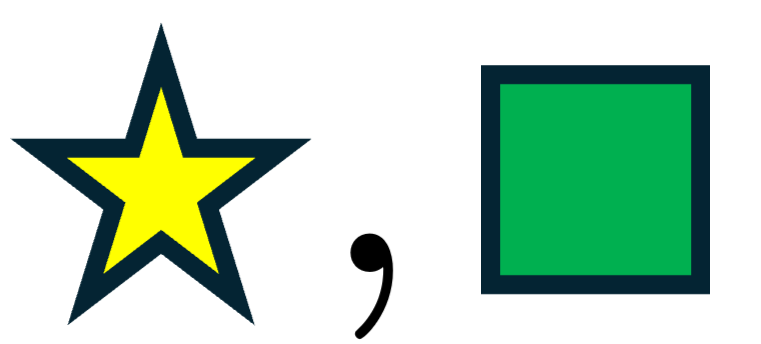} \\ \bottomrule
    \vspace{1mm}
    \end{tabular}
    \includegraphics[width=0.72\linewidth]{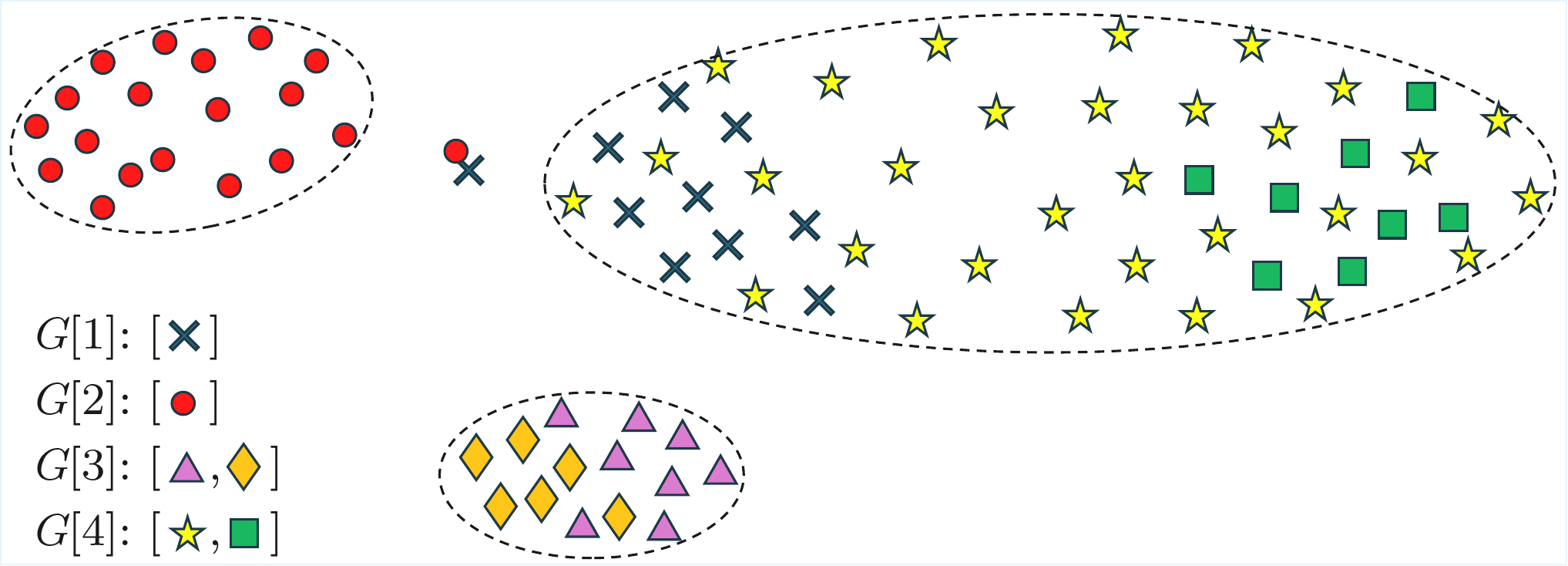}
    \caption{Different ways of measuring the distance between sets of clusters can lead to different aggregation results. In the table, $\text{\#tot} = \sum_{(i,a) \in G[r]}|P_i^a| \cdot \sum_{(j,b) \in G[s]} |P_j^b|$.}
    \label{fig:merging_example}
\end{figure}

Consider merging the clusters in Figure \ref{fig:merging_example} by iteratively merging the two sets of clusters with minimum distance. As shown in the attached table, the cluster of blue crosses can be aggregated with different clusters depending on the chosen definition of distance between two sets of clusters $G[r]$ and $G[s]$. In the figure, the dotted ellipses represent the true clustering solution $\mathcal{O}$. The first distance in the table is the minimum distance between any two points $x \in G[r], y \in G[s]$, and is not reliable in case of outliers. The second distance is the minimum over all the pairs of clusters $\bigl\{ (P_i^a, P_j^b) \ | \ P_i^a \in G[s],\ P_j^b \in G[s] \bigr\}$ of the average distance between the data points of the two clusters. The third distance is the average distance between all the pair of points $(x,y)$ such that $x \in G[r], y \in G[s]$. Both the second and third distance consider average distances, penalizing the cluster of yellow stars which contains several points distant from the cluster of blue crosses. The same happens using other distances inspired by common linkage criteria, such as median linkage, centroid linkage or Ward linkage \cite{ward1963hierarchical}, which are not show in the table.
The last distance in the table is minus the attractive force \eqref{eq:attractive_force_sets_of_clusters}, and for suitable values of $w$ provides the correct solution.
In particular, $w$ must not be excessively large, otherwise the attractive force of the red dot outlier may overpower all the other forces: if there are $N>0$ points exerting a force $\alpha>0$, and a single point exerting a force $\beta > \alpha$, if $w > \log_{\beta / \alpha}(N)$ then $N \alpha^w < \beta^w$.
In this example, another effective approach is to merge the two sets of clusters for which the increase in variance $\text{Var}(G[r] \cup G[s]) - \text{Var}(G[r]) - \text{Var}(G[s])$ is minimum. However, the minimum variance approach is not suitable for clusters with non-standard shapes, and in our simulations it proved inferior to the proposed approach.

\subsection{Algorithm description}

\algorithmstyle{ruled}
\begin{algorithm}
\linespread{1.35}\selectfont
\DontPrintSemicolon
\SetNoFillComment
\SetArgSty{textnormal}
\SetCommentSty{mycommentfont}
\tcp{Local clustering and probability density function fitting}
\For{each participant $i \in [n]$ in parallel}{
    labels $\leftarrow$ localClustering($X_i$) \par
    $\bigl\{ \hat{\theta}_i^a, \; \forall a \in [K_i] \bigr\} \leftarrow$ fitPDF($X_i$, labels)
    }
\tcp{Global aggregation at the server}
$ \bigl\{ P_i^a, \ \forall i \in [n], \ \forall a \in [K_i] \bigr\} \leftarrow $ generateProxyClusters($\bigl\{ \hat{\theta}_i^a,\ \forall i \in [n], \ \forall a \in [K_i]  \bigr\}$)\par
$G \leftarrow [ \ [(i,a)] \; \forall i \in [n], \ \forall a \in [K_i] \ ]$\par
\For{k in [|G| - 1]}{
    $H \leftarrow \left\{ f (G[r], G[s]) \; \forall r,s \in [|G|], \ s > r \right\}$\par
    $(r,s) \leftarrow \argmax(H)$\par
    $G[r] \leftarrow G[r] \cup G[s]$, and remove $G[s]$\par
    $Q_{|G|-k} \leftarrow G$ 
    }
Determine the optimal number of clusters $\hat{K}$.\par
Merge local clusters according to $Q_{\hat{K}}$.
\caption{One-shot Distributed Agglomerative Clustering with Masked Data}
\label{alg:one_shot}
\end{algorithm}

We describe each step of Algorithm \ref{alg:one_shot} in detail. In the local phase, participants run a clustering algorithm of their choice on their local dataset, individually and in parallel (line 2). This flexibility allows individual participants to select the most suitable clustering algorithm based on their computing resources.
Once they have converged to a local clustering solution, participants fit a probability distribution on each of their local clusters. In particular, participant $i$ fits a probability function on each local cluster, e.g. via maximum likelihood estimation, obtaining the set of parameters $\bigl\{ \hat{\theta}_i^a,\ \forall i \in [n], \ \forall a \in [K_i] \bigr\}$ where $K_i$ is the number of local clusters (line 3). The parametrized probability distributions are sent to the central server, which uses them to generate the synthetic dataset $\bigl\{ P_i^a, \ \forall i \in [n], \ \forall a \in [K_i] \bigr\}$, where $P_i^a$ is the proxy dataset associated with the $a$-th local cluster of participant $i$.

By using these synthetic representations of local clusters, the central server can aggregate local solutions and merge whole clusters instead of centroids. This approach may be better than simply comparing centroids, which neglects cluster covariance and is not robust if local data distributions are not uniform.
Since participants do not share raw data but only statistical distributions, the synthetic datasets protect anonymity at the level of single data points. Moreover, for some distributions such as the Dirichlet one, the probability density function is parametrized by a single $d_i$-dimensional vector, resulting in no communication overhead compared to transmitting the cluster centroid.

Proxy cluster are aggregated as described in Section \ref{sec:attractive_force}, updating a list $G$ whose elements specify which proxy clusters have been merged. More specifically, each element of $G$ is a list of pairs that have been merged, where each pair $(i,a)$ identifies a specific proxy cluster $P_i^a$. At the beginning of the global aggregation phase, all proxy clusters are disconnected. Merging is done iteratively, and continues until $G$ contains only one element, or $K$ elements if the number of clusters is given (line 6). At each iteration, we compute the force of attraction between all the sets of proxy clusters in $G$, and merge the two sets with the highest force of attraction (lines 7-9). If the optimal number of clusters $K$ is not known, it is convenient to store intermediate merging solutions, for example using additional variables (line 10) or creating a tree structure. By evaluating each of these intermediate solutions using a standard criterion such as \cite{rousseeuw1987silhouettes}, it is possible to estimate the optimal number of clusters and merge proxy clusters accordingly (lines 11-12).

\subsection{Algorithm analysis}

Below we discuss the correctness of Algorithm \ref{alg:one_shot} and under what conditions we can expect convergence to the optimal clustering solution $\mathcal{O}, \mathscr{o}$.
In the local phase of the algorithm, participants perform clustering on their local datasets using clustering algorithms of their choice. The analysis is identical to that in Section \ref{sec:alg_1_analysis_local_clustering}.
For simplicity, we assume that the local clustering solutions satisfy Assumption \ref{assumption:correct_local_overclustering}.

In the aggregation phase of the algorithm, the server compares and merges the local clusters of all participants. To ensure that the overlaps between the feature spaces of participants do not prevent necessary comparisons between local clusters, either Assumption \ref{assumption:graph} or Assumption \ref{assumption:graph_fully_connected} are needed. Under Assumption \ref{assumption:graph_fully_connected}, for any $a \in [K^\star]$ the feature spaces of any two participants $i,j \in \mathcal{V}^a$ overlap, allowing comparisons across all local clusters associated with the same true cluster. Under Assumption \ref{assumption:graph}, the graph $\mathcal{V}^a$ is connected $\forall a \in [K^\star]$ and data belonging to the same true cluster have identical distribution across participants. Therefore, local clusters from participants with disjoint feature spaces may be compared indirectly using the transitive property on a chain of local clusters in compatible feature spaces.

In Algorithm \ref{alg:one_shot} we are implicitly assuming that the distances between points in proxy datasets can be interchanged with the ones between points in the original local clusters. To ensure that this step is legitimate, we introduce Assumption \ref{assumption:for_algorithm2}, which is similar to Assumption \ref{assumption:for_methodA}. Since synthetic data points in proxy clusters do not have a direct relation with the non-masked dataset $\bar{X}$, we leverage suitable mappings between proxy clusters and the corresponding original clusters.
Let $\varphi_i^a (\cdot): P_i^a \to C_i^a$ map each synthetic data point in the proxy cluster $P_i^a$ to the nearest data point in the corresponding original cluster $C_i^a$:
\begin{equation} \label{eq:varphi}
    \varphi_i^a (x \in  P_i^a) = \argmin_{x' \in C_i^a} \phi(x, x').
\end{equation}
Let $\varphi(\cdot)$ be the union of all the maps $\varphi_i^a$, $\forall i \in \mathcal{V}^a,\ \forall a \in [K_i]$.

\begin{assumption} \label{assumption:for_algorithm2}
    Let $P_i^\square$ denote an arbitrary proxy cluster from participant $i$.
    $\forall a,b \in [K^\star],\ b \neq a$, and $\forall i \in \mathcal{V}^a$, $\forall j \in \mathcal{V}^a$, $\forall k \in \mathcal{V}^b$ such that $j,k \in \mathcal{N}(i) = \{ v \in [n] \ | \ \overleftrightsmallarrow{iv} \in \mathcal{E} \}$ it holds:
    \begin{equation} \label{eq:for_algorithm2}
        \begin{cases}
        \phi \bigl( \mho(\varphi(x_1)), \mho(\varphi(x_2)) \bigr) < \phi \bigl( \mho(\varphi(x_1)), \mho(\varphi(x_3)) \\
        \mho(\varphi(x_1)), \mho(\varphi(x_2)) \in \convexhull \bigl( \bar{X}^a \bigr) \\
        \mho(\varphi(x_3)) \in \convexhull \bigl( \bar{X}^b \bigr)
        \end{cases}
        \Rightarrow d_{ij}^\mathcal{P} (x_1, x_2) < d_{ik}^\mathcal{P} (x_1, x_3)
    \end{equation}
    $$\forall x_1 \in P_i^\square,\ \forall x_2 \in P_j^\square,\ \forall x_3 \in P_k^\square.$$
\end{assumption}

In Assumption \ref{assumption:for_algorithm2}, we use the map \eqref{eq:mho} to realize the mapping $\mho(\cdot): C_i^a \subseteq X_i \mapsto \bar{X}_i$, $\forall i \in [n]$, $\forall a \in [K^\star]$, providing the non-masked version of a generic data point. Therefore, the combined map
$$(\mho \circ \varphi)(\cdot): x \in P_i^a \mapsto \varphi(x) \in C_i^a \subseteq X_i \mapsto \mho(\varphi(x)) \in \bar{X}_i$$
maps a point in a proxy cluster to the nearest point in the corresponding original local cluster, and then to the non-masked version of the latter.
Using this mapping, we assume that calculating distances using proxy clusters is almost equivalent to using the original local clusters. Intuitively, this is true if proxy clusters are sufficiently similar to the original clusters they represent, which is verified if both the original and the proxy clusters contain a sufficient number of data points, as shown in Section \ref{sec:synthetic_dataset}. In practice, for computational efficiency the cardinality of proxy clusters is much smaller than that of the original local clusters they represent.

We summarize the combined effect of the various assumptions on the expected performance of Algorithm \ref{alg:one_shot}. Assumption \ref{assumption:correct_local_overclustering} guarantees that the local clustering solutions are correct up to overclustering, by this meaning that no local cluster contains data points belonging to multiple true clusters.
Assumption \ref{assumption:for_algorithm2} ensures that distances computed on synthetic data in different feature subspaces are reliable, allowing the iterative merging procedure to be to run on proxy clusters. Under this assumption, the two sets of proxy clusters for which the mutual attractive force is highest should represent the sets of original local clusters that are most similar to each other in the non-masked feature space.
As discussed in Section \ref{sec:attractive_force}, for suitable values of $w>1$ the iterative merging procedure is supposed to correctly aggregate clusters even when they are biased.
If either Assumption \ref{assumption:graph} or Assumption \ref{assumption:graph_fully_connected} is satisfied, the way data are distributed and masked among participants allows us to consider all available data points and compute the necessary attractive forces between proxy clusters. Overall, the global aggregation procedure is expected to merge the correct local clusters, leading to the optimal clustering solution $\mathcal{O}, \mathscr{o}$.

\subsection{Guarantees on the synthetic dataset} \label{sec:synthetic_dataset}

In Algorithm \ref{alg:one_shot} we fit probability distributions on the local clusters and transmit them to the central server. The server generates samples from these empirical distributions to obtain a synthetic dataset, which is then used to compute proxy distances and aggregate local clusters. This procedure allows to compare local clusters by taking into account their variance and probability function in addition to the cluster centroid, which alone may not be sufficient. However, for this approach to be beneficial we need to ensure that proxy clusters are sufficiently close to the original ones. In what follows, we compare a generic original cluster with the corresponding proxy cluster in terms of expected distance between their empirical data distributions. In particular, we consider the Wasserstein distance and total variation, two well-known statistical distances related to optimal transport.

Given a sequence of i.i.d. random variables $\{ x_i \in \mathbb{R}^d, \ i \in [N] \}$, the empirical probability measure is
$$\mu_N = \frac{1}{N} \sum_{i \in [N]} \delta(x_i)$$
where $\delta(\cdot)$ is the Dirac delta function. Let $X \in \mathbb{R}^{d \times N}$ be the original dataset containing data points belonging to a cluster, and $X_{\text{syn}} \in \mathbb{R}^{d \times M}$ the corresponding synthetic dataset.
Let $P$ be the true underlying distribution of $X$, and $P_{\hat{\theta}}$ be the fitted distribution with parameters $\hat{\theta}$ estimated via maximum likelihood estimation (MLE). Let $\mu_N$ be the empirical measure of $X$, and $\nu_M$ be the empirical measure of $X_{\text{syn}}$.

\subsubsection{Bound in Wasserstein-1 distance}
Given two probability distributions $\mu$ and $\nu$ on the same space, the Wasserstein distance tells the minimum cost of transforming one distribution into the other. In particular, by choosing the Euclidean distance, the Wasserstein-1 distance is defined as
$$
W_1(\mu, \nu) := \inf_{\gamma \in \Gamma(\mu, \nu)} \int_{\mathbb{R}^d \times \mathbb{R}^d} \norm{x - y} \, d\gamma(x, y)
$$
where $\Gamma(\mu, \nu)$ is the set of all the couplings of $\mu$ and $\nu$. In our case, if we imagine pairing up points from $X$ and $X_{\text{syn}}$ to minimize the sum of distances, $W_1(\mu, \nu)$ relates to this minimum sum. We can decompose the target distance using the triangle inequality:
$$ \mathbb{E}[W_1(\mu_N, \nu_M)] \le \mathbb{E}[W_1(\mu_N, P)] + \mathbb{E}[W_1(P, P_{\hat{\theta}})] + \mathbb{E}[W_1(P_{\hat{\theta}}, \nu_M)]$$

We now evaluate each of these terms separately.
\begin{itemize}
    \setlength\itemsep{0mm}
    \item Under fairly general conditions, e.g., $P$ has sufficiently many finite moments, it holds $\mathbb{E}[W_1(\mu_N, P)] \le O\left(N^{-1/d}\right)$ \cite{fournier2015rate, weed2019sharp}. Analogously, $\mathbb{E}[W_1(P_{\hat{\theta}}, \nu_M)] \le O\left(M^{-1/d}\right)$.
    \item The bound on the second term relies on the consistency and asymptotic normality of MLE. Under standard regularity assumptions, the MLE $\hat{\theta}$ converges to the true parameter $\theta^\star$ at a rate of $\sqrt{N}$. Specifically, $\sqrt{N}(\hat{\theta} - \theta^\star) \xrightarrow{D} \mathcal{N}(0, I(\theta^\star)^{-1})$, where $I(\theta^\star)$ is the Fisher information matrix \cite{lehmann1998theory}. Assuming Lipschitz continuity of the Wasserstein-1 distance with respect to the parameters, i.e. $ W_1(P_{\theta_1}, P_{\theta_2}) \leq L \norm{\theta_1 - \theta_2} $ for some $L>0$, we have $\mathbb{E}[W_1(P, P_{\hat{\theta}})] \le O\left(N^{-1/2}\right)$.
\end{itemize}
Overall, we get the combined bound
$$\mathbb{E}[W_1(\mu_N, \nu_M)] \le C_1 \left(\frac{1}{N^{1/d}}\right) + C_2 \left(\frac{1}{\sqrt{N}}\right) + C_3 \left(\frac{1}{M^{1/d}}\right)$$
where $C_1, C_2, C_3$ are constants that depend on the specific properties of the probability distribution $P$ and the dimensionality $d$. The $1/d$ exponent makes the rate of convergence degrade significantly in higher dimensions, showing the impact of the curse of dimensionality. However, for sufficiently large $N$ and $M$, the empirical distributions of the original and synthetic datasets are expected to be close, justifying our approach. To further reduce the deviation of proxy clusters from the original ones, participants could also share the bounding boxes associated with their local clusters, so that the data points generated at the central server are guaranteed to be within certain intervals.

\subsubsection{Bound in total variation for discrete distributions}

Total variation is another useful statistical distance between probability distributions. Given two discrete probability functions $\mu$ and $\nu$ on the same space, the total variation is half of their $L^1$ distance:
$$ \text{TV}(\mu, \nu) = \frac{1}{2} \sum_{x \in \mathcal{X}^d} |\mu(x) - \nu(x)| \in [0,1]. $$
Similarly as before, by applying the triangle inequality:
$$ \mathbb{E}[\text{TV}(\mu_N, \nu_M)] \le \mathbb{E}[\text{TV}(\mu_N, P)] + \mathbb{E}[\text{TV}(P, P_{\hat{\theta}})] + \mathbb{E}\text{TV}(P_{\hat{\theta}}, \nu_M)].$$
Let $b$ be the cardinality of the domain of the probability mass functions, i.e. the number of distinct values that $x$ can take.
For the first and last term, we have $\mathbb{E}[\text{TV}(\mu_N, P)] \leq \sqrt{b/N}/2$ and $\mathbb{E}[\text{TV}(P_{\hat{\theta}}, \nu_M)] \leq \sqrt{b/M}/2$ \cite{canonne2020short}. For the second term, the relationship between total variation and Kullback–Leibler divergence gives
$$\mathbb{E}[\text{TV}(P, P_{\hat{\theta}})]  \overset{(i)}{\leq} \mathbb{E}\left[\sqrt{\frac{1}{2} D_{KL}(P, P_{\hat{\theta}})}\right] \overset{(ii)}{\leq} \sqrt{ \mathbb{E}\left[ \frac{1}{2} D_{KL}(P, P_{\hat{\theta}})\right]} \overset{(iii)}{\leq} \sqrt{\frac{b-1}{N}}$$
where we used Pinsker's inequality \textit{(i)}, Cauchy-Schwarz \textit{(ii)}, and a result in \cite{roy2011bounds} \textit{(iii)}.
By summing all the terms, we get
$$ \mathbb{E}[\text{TV}(\mu_N, \nu_M)] \le \frac{1}{2} \sqrt{\frac{b}{N}} + \sqrt{\frac{b-1}{N}} + \frac{1}{2} \sqrt{\frac{b}{M}}.$$
Bounding the expected total variation in the case of continuous probability density functions requires using a different approach, as for absolutely continuous $P$ it holds $\text{TV}(\mu_N, P) = 1$ \cite{barron2002distribution}.

\subsection{Computational complexity}

Below we analyze the computational cost of Algorithm \ref{alg:one_shot} in terms of time complexity. For simplicity, assume all participants have $K_i = K$ clusters and $N$ data points in their local dataset.

In the first phase, each participant $i$ runs an arbitrary clustering algorithm on his local dataset, and fits a probability distribution on each cluster. The computational cost of local clustering depends on the chosen algorithm: for example, the time complexity per iteration of $K$-means is $\Theta(d_i K N)$ \cite{schutze2008introduction}, while the time complexity of agglomerative clustering with single-linkage is $O(d_i N^2)$ \cite{sibson1973slink}. Notably, each participant can choose a different algorithm based on the available computational resources.
The complexity of fitting probability distributions depends heavily on the type of distribution, number of parameters $\big| \hat{\theta} \big|$, and optimization method used. We consider the case where probability distributions are estimated via MLE and the likelihood is maximized via gradient ascent. Assuming each cluster contains $N/K$ data points, the complexity of a single gradient iteration for a single distribution is $O \bigl( d_i \big|\hat{\theta}\big| N / K \bigr)$. The complexity of fitting $K$ distributions, each requiring $I$ gradient iterations, is therefore $O \bigl( d_i \big|\hat{\theta}\big| N I \bigr)$.

In the second phase, the central server generates a proxy cluster for each local cluster. If each proxy cluster contains $M$ data points, generating the synthetic dataset is $O(nKM)$. Computing the attractive force between a pair of proxy clusters is $O(d M^2)$, and since there are $nK$ proxy clusters the number of pairs is $nK(nK-1)/2$. Aggregating the proxy clusters requires $nK-1$ iterations, where the $i$-th iteration involves at most $(nK-i)(nK-i-1)/2$ scalar comparisons to find the maximum attractive force and $O(nK-i-1)$ scalar sums to update the attractive forces. The total number of scalar comparisons is approximately equal to $\sum_{i=1}^{nK} i^2 = (nK)(nK+1)(2nK+1)/6$. Overall, the computational cost incurred by the central server is $O(d M^2 n^2 K^2 + n^3 K^3)$.
\section{Numerical experiments} \label{sec:experiments}

We implemented the proposed algorithms and tested them on three public datasets. Each dataset $\bar{X}$ is partitioned and masked according to either Assumption \ref{assumption:graph} or Assumption \ref{assumption:graph_fully_connected}. To satisfy Assumption \ref{assumption:graph}, we divide the $d$ features into $n$ partially overlapping sets, one for each participant, with $30\%$ overlap between two consecutive sets of features. To satisfy Assumption \ref{assumption:graph_fully_connected}, we select a small subset of features (approximately $10\%$ of the features) shared by all participants, and split the remaining features into $n$ disjoint sets, one for each participant. In both cases, we use feature importance analysis to ensure that the features given to each participant are sufficiently discriminative and allow for local clustering. All datasets include ground-truth labels and cluster memberships $\mathcal{O}$.
Information about each specific dataset is provided below.

\begin{itemize}
    \item Human Activity Recognition (HAR) \cite{human_activity_recognition_using_smartphones_240} is a dataset built from the recordings of 30 subjects performing activities of daily living while wearing inertial sensors on their waist. Each data point contains $d=561$ features describing a subject performing a certain activity during a time window. We apply clustering to distinguish between two activities performed by different subjects, setting $n = 30$ and $K_i = K = 2$ $\forall i \in [n]$. To satisfy Assumption \ref{assumption:graph} we distribute the available data randomly and evenly across participants, while to satisfy Assumption \ref{assumption:graph_fully_connected} each participant has data on a single subject.
    \item QMNIST \cite{qmnist-2019} is a dataset containing labeled images of handwritten digits, where each image consists of $d = 784$ pixels. We set $K=3$ and select the subset of digits $\{ 0, 3, 7 \}$ that can be distinguished with sufficient accuracy using standard clustering methods. Unlike the well-known MNIST dataset, QMNIST features extensive labels that allow for identification of different writers. To satisfy Assumption \ref{assumption:graph} we distribute the available data randomly and evenly across participants, while to satisfy Assumption \ref{assumption:graph_fully_connected} each participant has data from a single writer. In both cases we consider $n=10$ participants.
    \item DIM-sets \cite{ClusteringDatasets} is a collection of synthetic datasets for benchmarking clustering algorithms, with $K=16$ and multiple choices of $d \in [32, 1024]$. We distribute the data points belonging to each cluster among a random subset of participants, chosen according to their feature spaces. To satisfy Assumption \ref{assumption:graph} we distribute the data points randomly and evenly across the selected participants, while to satisfy Assumption \ref{assumption:graph_fully_connected} we distribute the data points sorted according to their first coordinate, so to introduce local bias. We choose $d=128$ features, $n=10$ participants, and $K_i < K$ $\forall i \in [n]$ local clusters.
\end{itemize}

\subsection{Testing the assumptions}

To assess the feasibility of the assumptions used in our analyses, for each dataset we check whether the following assumption is satisfied.

\begin{assumption} \label{assumption:experiments}
    $\forall a,b \in [K^\star],\ b \neq a$, and $\forall i \in \mathcal{V}^a$, $\forall j \in \mathcal{V}^a$, $\forall k \in \mathcal{V}^b$ such that $j,k \in \mathcal{N}(i) = \{ v \in [n] \ | \ \overleftrightsmallarrow{iv} \in \mathcal{E} \}$ it holds:
    \begin{equation}
        \phi (x_1, x_2) < \phi (x_1, x_3) \Rightarrow d_{ij}^{\bar{X}} (x_1, x_2) < d_{ik}^{\bar{X}} (x_1, x_3)
    \end{equation}
    $$\forall x_1 \in \bar{X}_i^a,\ \forall x_2 \in \bar{X}_j^a,\ \forall x_3 \in \bar{X}_k^b.$$
\end{assumption}

Assumption \ref{assumption:experiments} concerns the effect of data masking on the distances between data points. If Assumption \ref{assumption:experiments} is satisfied, we can expect Assumptions \ref{assumption:for_methodA} and \ref{assumption:for_algorithm2} to also be satisfied, since they are based on similar principles.
In all our experiments $\phi(x, x') = \norm{x - x'}$, the standard Euclidean distance. We repeat the evaluation multiple times for each dataset, considering different data partitions and feature masks. We test both the versions of the dataset that satisfy Assumption \ref{assumption:graph} and the ones that satisfy Assumption \ref{assumption:graph_fully_connected}. Since testing for all triplets of points is not feasible, we sample and test $1000$ triplets for each dataset configuration. The average satisfaction rate, shown in Table \ref{table:GCOD_satisfaction}, is very high in all cases, indicating that Assumption \ref{assumption:experiments} is a reasonable assumption and is often verified in practice.

\begin{table}[h]
    \centering
    \begin{tabular}{@{}c|cc@{}}
    \toprule
    Dataset & Assumption \ref{assumption:graph} & Assumption \ref{assumption:graph_fully_connected} \\ \midrule
    HAR     & $0.999$       & $0.999$     \\
    QMNIST  & $0.948$       & $0.976$    \\
    DIM-set$128$  & $1.000$      & $1.000$    \\ 
    \bottomrule
    \end{tabular}
    \vspace{2mm}
    \caption{Average satisfaction rate of Assumption \ref{assumption:experiments}.}
    \label{table:GCOD_satisfaction}
\end{table}

\subsection{Test of Algorithm \ref{alg:federated}}

We run Algorithm \ref{alg:federated} multiple times on each dataset, each time with a different random partition of the data points and features. All participants perform local clustering using $K$-means. We set $F=3$ federated iterations and a stepsize $\alpha=0.8$. We consider three evaluation metrics:
\begin{itemize}
    \item E1: correct aggregation of the local centroids in the initialization of the algorithm. E1 $\in [0, 1]$, where $1$ indicates perfect aggregation.
    \item E2: accuracy of the final clustering solution, computed with respect to the optimal clustering solution $\mathcal{O}$. The final cluster memberships are obtained using the final global centroids $G_c$. The optimal clustering solution $\mathcal{O}$ is given by the ground truth labels of the data points. For comparison, we also show the accuracy of centralized $K$-means applied on the original dataset $\bar{X}$, which includes all the features. Accuracies are expressed as percentages.
    \item E3: correctness of the final centroids. We first match the estimated global centroids in $G_c$ with the optimal centroids in $\mathscr{o}$, then we compute the cosine similarity and relative Euclidean distance for each pair of matched centroids, and finally we compute the average over the pairs. For example, the average Euclidean distance is computed as $ \mathbb{E}_a \bigl[ \norm{G_c[a] - o^a} / \norm{o^a} \bigr] $.
\end{itemize}
 
The results of our tests are contained in Tables \ref{table:results_alg_1_graph} and \ref{table:results_alg_1_graph_fully_connected}, which show the mean and standard deviation of the evaluation metrics across the various tests. Algorithm \ref{alg:federated} always correctly aggregates the local centroids and provides perfect or near-perfect estimates of the optimal centroids $\mathscr{o}$. The accuracy of the clustering solution is limited by the data distribution: in some cases, clusters are partially overlapping and perfect clustering is impossible. In particular, following feature masking, some samples of the QMNIST dataset are unrecognizable to the human eye. As a result, the local clustering solutions obtained by the participants contain several errors, resulting in imperfect centroids. Also in these cases, Algorithm \ref{alg:federated} provides clustering solutions whose accuracy is comparable to that of centralized $K$-means applied on the complete dataset $\bar{X}$, which includes all the features. By adopting either Method A or Method B based on whether Assumption \ref{assumption:graph} or \ref{assumption:graph_fully_connected} is satisfied, Algorithm \ref{alg:federated} algorithm effectively handles local datasets with non-uniform data distributions and participants with incompatible feature spaces.

\begin{table}[h]
    \centering
    \begin{tabular}{@{}c|ccc@{}}
    \toprule
    Dataset & E1 & \begin{tabular}[c]{@{}c@{}}E2\\ Alg. \ref{alg:federated} (Centralized $K$-means)\end{tabular} & \begin{tabular}[c]{@{}c@{}}E3\\ Cosine sim. $-$ Euclidean dist.\end{tabular} \\ \midrule
    HAR & $1.0 \pm 0.0$ & $99.87 \pm 0.04$ $(99.91)$ & $1.000 \pm 0.000 \ - \ 0.005 \pm 0.001$ \\
    QMNIST & $1.0 \pm 0.0$ & $92.16 \pm 1.28$ $(93.99)$ & $0.998 \pm 0.001 \ - \ 0.058 \pm 0.021$ \\
    DIM-set$128$ & $1.0 \pm 0.0$ & $100 \pm 0.00$ $(100)$ & $1.0 \pm 0.0 \ - \ 0.0 \pm 0.0$ \\ \bottomrule
    \end{tabular}
    \vspace{2mm}
    \caption{Empirical performance of Algorithm \ref{alg:federated} with Method A under Assumption \ref{assumption:graph}.}
    \label{table:results_alg_1_graph}
\end{table}

\begin{table}[h]
    \centering
    \begin{tabular}{@{}c|ccc@{}}
    \toprule
    Dataset & E1 & \begin{tabular}[c]{@{}c@{}}E2\\ Alg. \ref{alg:federated} (Centralized $K$-means)\end{tabular} & \begin{tabular}[c]{@{}c@{}}E3\\ Cosine sim. $-$ Euclidean dist.\end{tabular} \\ \midrule
    HAR & $1.0 \pm 0.0$ & $99.94 \pm 0.01$ $(99.90)$ & $1.000 \pm 0.000 \ - \ 0.002 \pm 0.001$ \\
    QMNIST & $1.0 \pm 0.0$ & $94.59 \pm 0.46$ $(96.66)$ & $0.998 \pm 0.001 \ - \ 0.059 \pm 0.009$ \\
    DIM-set$128$ & $1.0 \pm 0.0$ & $100 \pm 0.00$ $(100)$ & $1.0 \pm 0.0 \ - \ 0.0 \pm 0.0$ \\ \bottomrule
    \end{tabular}
    \vspace{2mm}
    \caption{Empirical performance of Algorithm \ref{alg:federated} with Method B under Assumption \ref{assumption:graph_fully_connected}.}
    \label{table:results_alg_1_graph_fully_connected}
\end{table}

\subsection{Test of Algorithm \ref{alg:one_shot}}

We run Algorithm \ref{alg:one_shot} multiple times on each dataset, considering different random partitions of the data points and features. All participants perform local clustering using $K$-means\texttt{++}, and fit multivariate Gaussian distributions on the resulting clusters. The server generates proxy clusters containing $30 \leq M \leq 100$ synthetic data points, and computes attractive forces using $w=2$. We evaluate the clustering performance using E3 and the following evaluation metrics:
\begin{itemize}
    \item E4: correct aggregation of the local proxy clusters. E4 $\in [0, 1]$, and $1$ indicates perfect aggregation.
    \item E5: accuracy of the final clustering solution, computed with respect to the optimal clustering solution $O$. This metric is analogous to E2, but here the final cluster labels and memberships are obtained from $Q_{\hat{K}}$.
\end{itemize}

\begin{table}[h!]
    \centering
    \begin{tabular}{@{}c|ccc@{}}
    \toprule
    Dataset & E4 & \begin{tabular}[c]{@{}c@{}}E5 \\ Alg. \ref{alg:one_shot} (Centralized $K$-means)\end{tabular} & \begin{tabular}[c]{@{}c@{}}E3\\ Cosine sim. $-$ Euclidean dist.\end{tabular} \\ \midrule
    HAR & $1.0 \pm 0.0$ & $99.87 \pm 0.04$ $(99.91)$ & $1.0 \pm 0.0 \ - \ 0.005 \pm 0.001$ \\
    QMNIST & $1.0 \pm 0.0$ & $91.37 \pm 1.35$ $(93.99)$ & $0.998 \pm 0.002 \ - \ 0.066 \pm 0.024$ \\
    DIM-set$128$ & $1.0 \pm 0.0$ & $100 \pm 0.00$ $(100)$ & $1.0 \pm 0.0 \ - \ 0.0 \pm 0.0$ \\ \bottomrule
    \end{tabular}
    \vspace{2mm}
    \caption{Empirical performance of Algorithm \ref{alg:one_shot} under Assumption \ref{assumption:graph}.}
    \label{table:results_alg_2_graph}
\end{table}

\begin{table}[h!]
    \centering
    \begin{tabular}{@{}c|ccc@{}}
    \toprule
    Dataset & E4 & \begin{tabular}[c]{@{}c@{}}E5 \\ Alg. \ref{alg:one_shot} (Centralized $K$-means)\end{tabular} & \begin{tabular}[c]{@{}c@{}}E3\\ Cosine sim. $-$ Euclidean dist.\end{tabular} \\ \midrule
    HAR & $1.0 \pm 0.0$ & $99.94 \pm 0.00$ $(99.90)$ & $1.0 \pm 0.0 \ - \ 0.002 \pm 0.001$ \\
    QMNIST & $1.0 \pm 0.0$ & $94.36 \pm 0.45$ $(96.66)$ & $0.998 \pm 0.001 \ - \ 0.062 \pm 0.011$ \\
    DIM-set$128$ & $1.0 \pm 0.0$ & $100 \pm 0.00$ $(100)$ & $1.0 \pm 0.0 \ - \ 0.0 \pm 0.0$ \\ \bottomrule
    \end{tabular}
    \vspace{2mm}
    \caption{Empirical performance of Algorithm \ref{alg:one_shot} under Assumption \ref{assumption:graph_fully_connected}.}
    \label{table:results_alg_2_graph_fully_connected}
\end{table}

Tables \ref{table:results_alg_2_graph} and \ref{table:results_alg_2_graph_fully_connected} show the mean and standard deviation of the evaluation metrics across the various tests. Table \ref{table:results_alg_2_graph} contains the results of the tests on the datasets that satisfy Assumption \ref{assumption:graph}, and is almost identical to Table \ref{table:results_alg_1_graph}. The only notable difference is the runtime: using $K$-means for local clustering in both algorithms, Algorithm \ref{alg:federated} is significantly faster than Algorithm \ref{alg:one_shot}. Table \ref{table:results_alg_2_graph} presents the tests on the datasets that satisfy Assumption \ref{assumption:graph_fully_connected}, and also in this case the values of the evaluation metrics are similar. This demonstrates the effectiveness of Algorithm \ref{alg:one_shot} even when clusters have moderately different data distributions for each participant.
\section{Conclusions} \label{sec:conclusions}

In this work we introduce and formalize the problem of distributed clustering in partially overlapping feature spaces: each participant observes only a fixed subset of features, with some overlap between the sets of features observed by different participants. We propose, analyze and test two distributed algorithms that address the above problem. Algorithm \ref{alg:federated} is a federated algorithm that iteratively updates a set of global centroids, and is available in two variants suitable for different problem settings. Algorithm \ref{alg:one_shot} is a one-shot algorithm: participants share the parameterizations of their local clusters, which are used by the server to generate and merge synthetic clusters. The proposed algorithms effectively handle the challenges of distributed private data and partially-observable features, and offer tunable computational cost and communication efficiency.

By addressing a novel clustering problem, this work opens up interesting possibilities for future research. Different assumptions can be made about the data and on how they are partitioned among participants. For example, some data types may require the use of kernels, autoencoders, or aggregation techniques based on different principles. Similarly, different assumptions can be made about how features are distributed among participants and about the relationship between features. These assumption may lead to different algorithms or to more rigorous theoretical guarantees. Furthermore, it is possible to develop fully-distributed algorithms that do not require a central server.

\bibliographystyle{ieeetr}
\bibliography{references}

\begin{thebibliography}{10}

\bibitem{rubin1976inference}
D.~B. Rubin, ``Inference and missing data,'' {\em Biometrika}, vol.~63, no.~3, pp.~581--592, 1976.

\bibitem{emmanuel2021survey}
T.~Emmanuel, T.~Maupong, D.~Mpoeleng, T.~Semong, B.~Mphago, and O.~Tabona, ``A survey on missing data in machine learning,'' {\em Journal of Big data}, vol.~8, no.~1, p.~140, 2021.

\bibitem{fan2020polynomial}
J.~Fan, Y.~Zhang, and M.~Udell, ``Polynomial matrix completion for missing data imputation and transductive learning,'' in {\em Proceedings of the AAAI Conference on Artificial Intelligence}, vol.~34, pp.~3842--3849, 2020.

\bibitem{sun2023deep}
Y.~Sun, J.~Li, Y.~Xu, T.~Zhang, and X.~Wang, ``Deep learning versus conventional methods for missing data imputation: A review and comparative study,'' {\em Expert Systems with Applications}, vol.~227, p.~120201, 2023.

\bibitem{wang2019k}
S.~Wang, M.~Li, N.~Hu, E.~Zhu, J.~Hu, X.~Liu, and J.~Yin, ``K-means clustering with incomplete data,'' {\em IEEE Access}, vol.~7, pp.~69162--69171, 2019.

\bibitem{zhang2021gaussian}
Y.~Zhang, M.~Li, S.~Wang, S.~Dai, L.~Luo, E.~Zhu, H.~Xu, X.~Zhu, C.~Yao, and H.~Zhou, ``Gaussian mixture model clustering with incomplete data,'' {\em ACM Transactions on Multimedia Computing, Communications, and Applications (TOMM)}, vol.~17, no.~1s, pp.~1--14, 2021.

\bibitem{chi2016k}
J.~T. Chi, E.~C. Chi, and R.~G. Baraniuk, ``k-pod: A method for k-means clustering of missing data,'' {\em The American Statistician}, vol.~70, no.~1, pp.~91--99, 2016.

\bibitem{wagstaff2004clustering}
K.~Wagstaff, ``Clustering with missing values: No imputation required,'' in {\em Classification, Clustering, and Data Mining Applications: Proceedings of the Meeting of the International Federation of Classification Societies (IFCS), Illinois Institute of Technology, Chicago, 15--18 July 2004}, pp.~649--658, Springer, 2004.

\bibitem{singh2013comparative}
D.~Singh and A.~Gosain, ``A comparative analysis of distributed clustering algorithms: A survey,'' in {\em 2013 International Symposium on Computational and Business Intelligence}, pp.~165--169, IEEE, 2013.

\bibitem{mcmahan2017communication}
B.~McMahan, E.~Moore, D.~Ramage, S.~Hampson, and B.~A. y~Arcas, ``Communication-efficient learning of deep networks from decentralized data,'' in {\em Artificial intelligence and statistics}, pp.~1273--1282, PMLR, 2017.

\bibitem{garst2024federated}
S.~Garst and M.~Reinders, ``Federated k-means clustering,'' in {\em International Conference on Pattern Recognition}, pp.~107--122, Springer, 2024.

\bibitem{zhu2023f3km}
S.~Zhu, Q.~Xu, J.~Zeng, S.~Wang, Y.~Sun, Z.~Yang, C.~Yang, and Z.~Peng, ``F3km: Federated, fair, and fast k-means,'' {\em Proceedings of the ACM on Management of Data}, vol.~1, no.~4, pp.~1--25, 2023.

\bibitem{zhou2022memory}
X.~Zhou and X.~Wang, ``Memory and communication efficient federated kernel k-means,'' {\em IEEE Transactions on Neural Networks and Learning Systems}, vol.~35, no.~5, pp.~7114--7125, 2022.

\bibitem{ngo2023federated}
H.~Ngo, H.~Fang, J.~Rumbut, and H.~Wang, ``Federated fuzzy clustering for decentralized incomplete longitudinal behavioral data,'' {\em IEEE internet of things journal}, vol.~11, no.~8, pp.~14657--14670, 2023.

\bibitem{li2025vertical}
Y.~Li, X.~Hu, S.~Yu, W.~Ding, W.~Pedrycz, Y.~C. Kiat, and Z.~Liu, ``A vertical federated multi-view fuzzy clustering method for incomplete data,'' {\em IEEE Transactions on Fuzzy Systems}, 2025.

\bibitem{yan2024federated}
X.~Yan, Z.~Wang, and Y.~Jin, ``Federated incomplete multi-view clustering with heterogeneous graph neural networks,'' in {\em International Workshop on Trustworthy Federated Learning}, pp.~61--76, Springer, 2024.

\bibitem{chen2023federated}
X.~Chen, J.~Xu, Y.~Ren, X.~Pu, C.~Zhu, X.~Zhu, Z.~Hao, and L.~He, ``Federated deep multi-view clustering with global self-supervision,'' in {\em Proceedings of the 31st ACM International Conference on Multimedia}, pp.~3498--3506, 2023.

\bibitem{chen2024bridging}
X.~Chen, Y.~Ren, J.~Xu, F.~Lin, X.~Pu, and Y.~Yang, ``{Bridging gaps: Federated multi-view clustering in heterogeneous hybrid views},'' {\em Advances in Neural Information Processing Systems}, vol.~37, pp.~37020--37049, 2024.

\bibitem{ren2024novel}
Y.~Ren, X.~Chen, J.~Xu, J.~Pu, Y.~Huang, X.~Pu, C.~Zhu, X.~Zhu, Z.~Hao, and L.~He, ``A novel federated multi-view clustering method for unaligned and incomplete data fusion,'' {\em Information Fusion}, vol.~108, p.~102357, 2024.

\bibitem{de2019isotropy}
S.~De~Iaco, D.~Posa, C.~Cappello, and S.~Maggio, ``Isotropy, symmetry, separability and strict positive definiteness for covariance functions: a critical review,'' {\em Spatial statistics}, vol.~29, pp.~89--108, 2019.

\bibitem{arbia2008class}
G.~Arbia, G.~Espa, and D.~Quah, ``A class of spatial econometric methods in the empirical analysis of clusters of firms in the space,'' {\em Empirical Economics}, vol.~34, no.~1, pp.~81--103, 2008.

\bibitem{johnson1984extensions}
W.~B. Johnson, J.~Lindenstrauss, {\em et~al.}, ``Extensions of lipschitz mappings into a hilbert space,'' {\em Contemporary mathematics}, vol.~26, no.~189-206, p.~1, 1984.

\bibitem{li2022binary}
W.-Y. Li and S.-Z. Zhang, ``Binary random projections with controllable sparsity patterns,'' {\em Journal of the Operations Research Society of China}, vol.~10, no.~3, pp.~507--528, 2022.

\bibitem{arthur2006k}
D.~Arthur and S.~Vassilvitskii, ``k-means++: The advantages of careful seeding,'' tech. rep., Stanford, 2006.

\bibitem{yang2024greedy}
K.~Yang, M.~Mohammadi~Amiri, and S.~R. Kulkarni, ``Greedy centroid initialization for federated k-means,'' {\em Knowledge and Information Systems}, vol.~66, no.~6, pp.~3393--3425, 2024.

\bibitem{rousseeuw1987silhouettes}
P.~J. Rousseeuw, ``Silhouettes: a graphical aid to the interpretation and validation of cluster analysis,'' {\em Journal of computational and applied mathematics}, vol.~20, pp.~53--65, 1987.

\bibitem{brandao2021efficient}
A.~Brand{\~a}o, R.~Mendes, and J.~P. Vilela, ``Efficient privacy preserving distributed k-means for non-iid data,'' in {\em International Symposium on Intelligent Data Analysis}, pp.~439--451, Springer, 2021.

\bibitem{holzer2023dynamically}
P.~Holzer, T.~Jacob, and S.~Kavane, ``Dynamically weighted federated k-means,'' {\em arXiv preprint arXiv:2310.14858}, 2023.

\bibitem{kuhn1955hungarian}
H.~W. Kuhn, ``The hungarian method for the assignment problem,'' {\em Naval research logistics quarterly}, vol.~2, no.~1-2, pp.~83--97, 1955.

\bibitem{ester1996density}
M.~Ester, H.-P. Kriegel, J.~Sander, X.~Xu, {\em et~al.}, ``A density-based algorithm for discovering clusters in large spatial databases with noise,'' in {\em kdd}, vol.~96, pp.~226--231, 1996.

\bibitem{schutze2008introduction}
H.~Sch{\"u}tze, C.~Manning, and P.~Raghavan, {\em Introduction to information retrieval}.
\newblock Cambridge University Press, 2008.

\bibitem{edmonds1972theoretical}
J.~Edmonds and R.~M. Karp, ``Theoretical improvements in algorithmic efficiency for network flow problems,'' {\em Journal of the ACM (JACM)}, vol.~19, no.~2, pp.~248--264, 1972.

\bibitem{ward1963hierarchical}
J.~H. Ward~Jr, ``Hierarchical grouping to optimize an objective function,'' {\em Journal of the American statistical association}, vol.~58, no.~301, pp.~236--244, 1963.

\bibitem{fournier2015rate}
N.~Fournier and A.~Guillin, ``On the rate of convergence in wasserstein distance of the empirical measure,'' {\em Probability theory and related fields}, vol.~162, no.~3, pp.~707--738, 2015.

\bibitem{weed2019sharp}
J.~Weed and F.~Bach, ``Sharp asymptotic and finite-sample rates of convergence of empirical measures in wasserstein distance,'' {\em Bernoulli}, vol.~25, no.~4A, pp.~2620--2648, 2019.

\bibitem{lehmann1998theory}
E.~L. Lehmann and G.~Casella, {\em Theory of point estimation}.
\newblock Springer, 1998.

\bibitem{canonne2020short}
C.~L. Canonne, ``A short note on learning discrete distributions,'' {\em arXiv preprint arXiv:2002.11457}, 2020.

\bibitem{roy2011bounds}
B.~Roy, ``Bounds on the expected entropy and kl-divergence of sampled multinomial distributions,'' {\em Unpublished manuscript, MIT}, 2011.

\bibitem{barron2002distribution}
A.~R. Barron, L.~Gyorfi, and E.~C. van~der Meulen, ``Distribution estimation consistent in total variation and in two types of information divergence,'' {\em IEEE transactions on Information Theory}, vol.~38, no.~5, pp.~1437--1454, 2002.

\bibitem{sibson1973slink}
R.~Sibson, ``Slink: an optimally efficient algorithm for the single-link cluster method,'' {\em The computer journal}, vol.~16, no.~1, pp.~30--34, 1973.

\bibitem{human_activity_recognition_using_smartphones_240}
J.~Reyes-Ortiz, D.~Anguita, A.~Ghio, L.~Oneto, and X.~Parra, ``{Human Activity Recognition Using Smartphones}.'' UCI Machine Learning Repository, 2013.
\newblock {DOI}: https://doi.org/10.24432/C54S4K.

\bibitem{qmnist-2019}
Y.~Chhavi and B.~L\'{e}on, ``Cold case: The lost mnist digits,'' in {\em Advances in Neural Information Processing Systems 32}, Curran Associates, Inc., 2019.

\bibitem{ClusteringDatasets}
P.~Fr\"anti and S.~Sieranoja, ``K-means properties on six clustering benchmark datasets,'' 2018.

\end{thebibliography}



\end{document}